\documentclass[aps,prl,onecolumn,superscriptaddress,preprintnumbers,showpacs]{revtex4}
\usepackage{psfrag,epsfig,amsmath,amssymb}

\def\a{\alpha} 
\def\b{\beta}

\def\ga{\gamma}

\def\m{\mu}
\def\n{\nu}
\def\r{\rho}
\def\s{\sigma}

\def\th{\theta}

\def\g5{\gamma_{5}}

\def\idx{\int\!\! d^4\!x}

 \def\Dirac{{D\mkern-12mu/}\,}

\begin{document}

\psfrag{ep}{$\epsilon(P)$}
\psfrag{Q}{$q$}
\psfrag{Qb}{$\bar{q}$}
\psfrag{t}{$\theta$}
\psfrag{ek1}{$\epsilon(k_1)$}
\psfrag{ek2}{$\epsilon(k_2)$}
\psfrag{g,Z}{$\gamma,Z$}
\psfrag{p}{$p$}
\psfrag{q}{$k$}
\psfrag{q1}{$k_1$}
\psfrag{q2}{$k_2$}
\psfrag{q3}{$k_3$}
\psfrag{Amu}{$A_\m$}
\psfrag{Zmu}{$Z_\m$}
\psfrag{Anu}{$A_\n$}
\psfrag{Arho}{$A_\r$}
\psfrag{Phib}{${\psi_\b}$}
\psfrag{Phica}{$\overline{\psi}_\a$}
\psfrag{xx}{$K_{Z\ga\ga}$}
\psfrag{yy}{$K_{\ga\ga\ga}$}
\psfrag{z1}{$BR^{nmNCSM}_{[J/\Psi\rightarrow\ga\ga]}$}
\psfrag{z2}{$BR^{nmNCSM}_{[\Upsilon\rightarrow\ga\ga]}$}
\title
{Noncommutative fermions and quarkonia decays}

\author{Carlos Tamarit}
\affiliation{Departamento de F\'{\i}sica Te\'orica I, Facultad de Ciencias F\'{\i}sicas, 
Universidad Complutense de Madrid, 28040 Madrid, Spain }
\author{Josip Trampeti\' c}
\affiliation{Theoretical Physics Division,
Rudjer Bo\v{s}kovi\'{c} Institute, P.O.Box 180,
10002 Zagreb, Croatia}
\date{\today}

\begin{abstract}
The recent introduction of a deformed non-minimal version of the noncommutative Standard Model
in the enveloping-algebra approach, having a one-loop renormalisable gauge sector
involving a higher order gauge term, motivates us to consider the possibility of extending 
the fermion sector with additional deformations, i.e. higher order fermionic terms. Since the renormalisability 
properties of the fermion sector of the model are not yet fully known, we work with an effective fermion 
lagrangian which includes noncommutative higher order terms involving a contraction with 
the noncommutative $\theta$ tensor aside from the star products, 
so that these terms annihilate in the commutative limit. 
Some of these terms violate CPT in the weak sector, and some violate CP in the strong and hypercharge sectors.
We apply this framework to the reevaluation
of the  decay rates of quarkonia ($\overline{q}q_{1} = J/\psi,\, \Upsilon$) into two photons. 
These decays, which are forbidden in the ordinary Standard Model, had been previously studied as  
possible signals for noncommutativity, but not in the framework of the better behaved deformed 
non-minimal version of the noncommutative Standard Model. 
Weak CPT or strong-hypercharge CP violating interactions do not contribute to the result. 
If the parameters of the model take natural values, for the vast majority of configurations 
the resulting branching ratios are
enhanced with respect to their values in the minimal version of the noncommutative Standard Model. 
Also for more than half of the parameter space, 
the rates are larger than the maximal 
rates that were calculated in the undeformed version of the non-minimal noncommutative Standard Model. Tuning the dimensionless parameters the predicted branching ratios can fall within the current experimental bounds.
\end{abstract}
\pacs{11.10.Nx, 12.38.-t, 12.39Dc, 12.39.-x, 14.20-c}
\maketitle

\section{Introduction}

 The Standard Model (SM) of particle physics and the theory of gravity
describe very well, as far as we know today, all physical phenomena 
from cosmological processes to the properties of subnuclear structures. 
Nevertheless, at extreme energies and/or
very short distances -at the Planck scale- this theories fail to be compatible, 
which motivates the study of
modified or alternative space-time structures
that could help to solve the above mentioned
difficulties or at least shed some light on them. 
These modified space-time structures arise in such settings as
the quantised coordinates  in string theory
or in the general framework of deformation quantisation.
The idea of noncommutative (NC) space-time, which can be realised in both of the above settings, 
has recently found more and more interest. In this paper we deal with noncommutative theories 
defined by means of the enveloping algebra approach, which allows to define gauge theories with 
arbitrary gauge groups, in particular that of the Standard Model. 
The research on these theories so far has successfully dealt with some 
theoretical and also phenomenological aspects, 
which might allow the confrontation of the theory with experiments.

On the theoretical side, in the enveloping algebra approach developed in 
\cite{Madore:2000en} following  the ideas of the seminal paper \cite{Seiberg:1999vs}, 
we can emphasise the construction of noncommutative minimal and non-minimal versions of 
the Standard Model (mNCSM and nmNCSM) \cite{Calmet:2001na,Behr:2002wx,Melic:2005fm}, and
GUTS \cite{Aschieri:2002mc}. Of particular interest is the deformed version of the  nmNCSM 
with renormalisable pure gauge interactions discussed in refs.~\cite{Buric:2006wm,Latas:2007eu}, 
which involves an extension of the pure noncommutative gauge lagrangian ${\widehat F}\star {\widehat F}$ 
with a deformation or higher order term. These noncommutative extensions 
of the SM are anomaly free 
\cite{Martin:2002nr,Brandt:2003fx}. In contrast with the gauge sector, 
the renormalisability of the fermion sector has not yet been completely addressed, 
despite some encouraging partial results \cite{Buric:2007ix}. 
Nevertheless, the results in the pure gauge sector motivate to consider 
the possible effects of introducing higher order fermionic terms;
in this paper we will initiate the study of an extended fermion sector by introducing 
a general class of the fermionic deformations in the noncommutative action of the nmNCSM 
and obtaining some of the corresponding Feynman rules.

On the other hand, on the phenomenological side, it is known that noncommutative field theories 
can predict non-zero rates for processes which are forbidden in 
the Standard Model due to the Lorentz invariance and Bose symmetry (Landau-Pomeranchuk-Yang 
or LPY theorem). These new effects follow from the violation of Lorentz invariance in the presence of 
noncommutativity, and the observation of these forbidden decays could be taken as a signal for it. 
Some of the SM forbidden processes that have been studied, at the level of tree diagrams, 
include $Z\rightarrow\ga\ga/gg$ \cite{Behr:2002wx}; 
$J/\Psi,\Upsilon\rightarrow\ga\ga$ \cite{Melic:2005hb} and $K\rightarrow\pi\ga$ \cite{Melic:2005su}. 
For other phenomenological studies, including limits on the noncommutative scale
$\Lambda_{\rm NC}$, see \cite{Alboteanu:2006hh}.
The recent introduction of the  deformed nmNCSM calls for a reevaluation of the results, 
which so far has  only been done for the $Z\rightarrow\ga\ga$ process \cite{Buric:2007qx}. 
In this paper we will apply our framework of a deformed fermionic sector to study 
the quarkonia decays into two photons $J/\Psi,\Upsilon\rightarrow\ga\ga$. 
There are other processes involving quarkonia in which noncommutativity induces rare decays. 
An example is the decay into two photons of quarkonia polarised in some direction, say the $x^3$ axis. 
In such polarised rate the contributions proportional to the third components $(E^3_{\theta})^2$ and $(B^3_{\theta})^2$ 
should be enhanced by a large factor, similarly as in the $Z \rightarrow \gamma\gamma$ case \cite{Buric:2007qx}.
In addition, one could study decays of quarkonia in other gauge bosons.
In general, we expect that any quark-antiquark state with 
the same quantum numbers as those of $J/\Psi$ and $\Upsilon$ ($I^G (J^{PC}) = 0^- (1^{--}))$ 
will decay into two gauge bosons through noncommutative interactions.  
We restrict ourselves to quarkonia decays into two photons because their detection could in principle be achieved 
within  very high-resolution calorimeters by applying more stringent conditions in the selection of the photon candidates
when searching for $\gamma\gamma$ events. 
We could also have decays into a pair of gluons, 
but these would hadronise into hadron jets, whose detection would be much more problematic due to a lack 
of localisation of the jets or interference with other signals. 
Of course there could be also decays into two Z bosons, but, on the one hand, 
they would not be produced for quarkonia at rest due to the heaviness of the Z bosons, and, on the other, 
these would rapidly decay into other particles.

The paper is organised as follows. 
First, we give some more detailed fundamental and phenomenological motivations for our work, 
and we introduce the theoretical framework.  Next we introduce the action of 
the renormalisable pure gauge sector of deformed nmNCSM, after which we deal with the matter sector, 
extending the fermion action 
with higher order deformations and examining their C,P,T transformation properties. 
Following this, we study whether field redefinitions in 
the fermion sector can help to get rid of some of the extra terms introduced. 
Finally we give the Feynman rules relevant to the calculation of the quarkonia 
decay rates and we present our results, which are then discussed.

\section{Motivation and framework}

Our main goal in this paper is, following the recent introduction of 
the deformed version of the gauge sector of the noncommutative Standard Model (NCSM),
to consistently define the action of an extended fermion sector of the nmNCSM from 
an effective-theory point of view, by adding higher order terms, and derive relevant Feynman rules. 
This is certainly important for  future investigations about the renormalisability 
properties of entire NCSM, which are interesting in their own right 
and could introduce constraints on our ``effective'' deformed fermion sector.
The second goal of this paper is to apply this framework to the calculation of 
the decay rates of the simplest processes which are forbidden in the ordinary SM, 
that is the C symmetry violating decay of quarkonia into two photons, 
($\overline{q}q_{1} = J/\psi,\, \Upsilon$). 
These processes are of theoretical interest because their tree-level contributions come from 
two different types of diagrams, these being s-channel gauge bosons exchanges 
--involving interactions from the pure gauge sector of the NCSM-- and
t-channel quark exchanges --involving interactions from the fermion sector. 
Of course, it should not be forgotten that 
this proposed processes are also  important because they violate the LPY theorem that holds in the ordinary SM, 
and thus their hypothetical detection could be taken as a possible signal of noncommutativity.

The first construction of the NCSM was undertaken in ref.~\cite{Calmet:2001na}, 
where it was already noted that there was an ambiguity in the choice of traces for the gauge kinetic terms, 
leading to a minimal version (with traces taken in the adjoint representation), and a non-minimal version 
including traces over the representations of all the massive particle multiplets 
charged under any of the gauge groups \cite{Behr:2002wx}.
The  interaction vertices and Feynman rules of 
these models were further analysed in refs.~\cite{Melic:2005fm}.  
As was said, some of the processes studied in this framework include 
$Z\rightarrow\ga\ga/gg$ \cite{Behr:2002wx,Buric:2007qx}; 
$J/\Psi,\Upsilon\rightarrow\ga\ga$ \cite{Melic:2005hb}, $K\rightarrow\pi\ga$ \cite{Melic:2005su}
and $\gamma_{plasmon}\rightarrow \bar\nu \nu$ \cite{Schupp:2002up}. 

At one loop order, the investigations so far were concerned with renormalisability properties
\cite{Buric:2006wm,Buric:2007qx,Latas:2007eu}; 
remarkably, it was found that the pure gauge interactions of the nmNCSM could be rendered one-loop 
renormalisable at first order in the noncommutativity parameters $\th^{\m\n}$ by adding an extra 
deformation term to the lagrangian involving only gauge fields and their derivatives contracted with one 
$\th^{\m\n}$. This term was introduced at 
the level of the noncommutative action \cite{Buric:2007qx,Latas:2007eu}, 
yielding  an extended version of the nmNCSM \cite{Buric:2006wm}. 
This result was reached by considering  only gauge 
field contributions to the loop integrals. When matter fields are included, 
 the results of ref.~\cite{Martin:2007wv} show that, when computing the matter contributions to 
 the one-loop diagrams with external gauge fields in a generic noncommutative gauge theory with Dirac 
 fermions or complex scalars, the divergences can be absorbed in the bare lagrangian whenever 
 the representations of the matter fields are included in the choice of representations for 
 the traces in the pure gauge terms of the action.

This result motivates the following comments. First, it suggests that there is virtually 
no hope of getting renormalisability of the gauge sector 
in the mNCSM case when Dirac fermions run in the loops, 
since the model only involves traces in the adjoint representation of the gauge fields. 
However, in the pure gauge sector of the nmNCSM \cite{Behr:2002wx,Melic:2005fm} 
the traces include the representations of the
matter fields, so that one could say that it is likely that the nmNCSM gauge sector will
still be renormalisable if one includes the effects of the matter
fields in the loops. This motivates even more to favour the use of the nmNCSM, 
and in particular its deformed version, over the mNCSM.

Regarding the renormalisability properties of the nmNCSM, 
we would like to make some clarifying observation.
The NC SU(N) pure gauge theory \cite{Latas:2007eu} showed renormalisability for 
two choices of the free deformation parameter $a$, associated with the higher order term, 
$S^H_g(a)$, that is present in the deformed nmNCSM: $a=1$ and $a=3$. 
However, this result cannot be extrapolated directly to the NCSM, because
the gauge fields mix after the SW map and it is not a sum of NC SU(N)
theories. In fact,  the gauge sector of the nmNCSM is
{\it only renormalisable} for $a=3$, even more, 
the one-loop {\it quantum corrections are finite} for $a=3$.

The above arguments clearly favours the nmNCSM over the mNCSM if one wants to have a better behaved theory, 
where the quantum corrections are more  under control even when working in an effective theory approach. 
In particular, predictions of the theory involving pure gauge boson interactions  will be more 
robust under changes of scale.

All these new results call for a reevaluation, in the framework of the deformed nmNCSM, of 
the possible signals for noncommutativity that were commented upon before. As was said before, 
this has already been done in ref.~\cite{Buric:2007qx} for the $Z\rightarrow\ga\ga$ decay. 
In this paper we will focus on the disintegration $J/\Psi,\Upsilon\rightarrow\ga\ga$. 
Now, since the renormalisability properties of the fermion sector of deformed nmNCSM 
are essentially unknown, despite promising results for some diagrams involving chiral 
fermions \cite{Buric:2007ix}, and since the results for the pure gauge sector make us expect 
that a renormalisable fermion sector could only be achieved by adding deformed fermionic 
contributions to the action,
it makes sense to treat this sector effectively and consider, aside from the fermion lagrangian 
employed in both mNCSM and nmNCSM, higher order contributions involving contractions with 
$\theta^{\m\n}$ outside the star product and compatible with the noncommutative gauge symmetry. 
We will demand that these deformed fermionic contributions do not introduce further 
violations of parity than those coming from the ordinary $SU(2)_L$ gauge fields; also, 
we will require that they do not alter the tree level fermion 2-point function.
This modification of the propagator could be achieved, for example, with terms like 
$i\bar{\hat\psi}\theta^{\a\b}\ga_\b\{\hat D^2,\hat D_\a\}\star\hat\psi$. The problem is that, 
when computing S-matrix elements, one ordinarily uses the LSZ formalism, which is constructed by 
using Lorentz invariance and  implies that the S-matrix elements are obtained by identifying 
the poles in momentum space of some Green functions. This happen at points satisfying 
the Lorentz invariant constraint $p^2=m^2$ for some $m$.
In noncommutative spacetime Lorentz invariance is broken, which means that the poles of 
the Green functions need not satisfy the constraint $p^2=m^2$; a $\theta$-dependence may 
appear and then one should be more careful when computing S-matrix elements. 
An alternative definition is needed, perhaps along the lines of the work done 
in ref.~\cite{Grosse:2008dk} for noncommutative theories formulated without SW maps. Nevertheless, 
if the tree-level propagator still has a pole of the ordinary type, we expect that the usual 
way of deriving matrix elements will be valid to some approximation. Here we pretend 
to compute  S-matrix elements at tree level only.

\section{Pure gauge sector}

\subsection{Renormalisable pure gauge sector action}

\begin{figure}[h]
\begin{center}
\hspace{-1cm}
\includegraphics[scale=0.75]{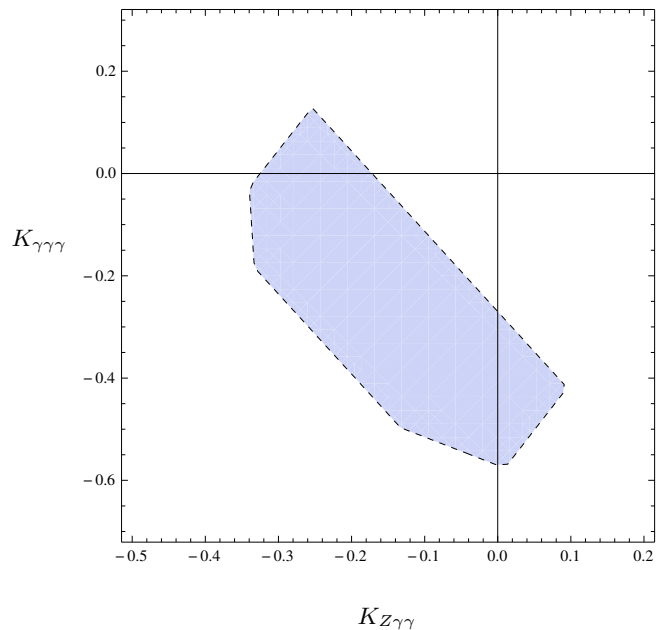}
\end{center}
\caption{Allowed values for the couplings $K_{Z\ga\ga}$ and $K_{\ga\ga\ga}$ in the nmNCSM.}
\label{f:0}
\end{figure}

The pure gauge part of the deformed nmNCSM is given by \cite{Buric:2007qx}:
\begin{align}
 S_g &= S^{min}_g + S^H_g(a),
 \nonumber\\
 S_g^{min} &=- \frac{1}{2}{\rm Tr}\int d^4x
\widehat F_{\mu\nu}\star\widehat F^{\mu\nu},\, &
S_g^H(a) &=\frac{a-1}{4}{\rm Tr}\int d^4xh \theta^{\mu\nu}\star
\widehat F_{\mu\nu}\star\widehat F_{\rho\sigma}\star\widehat F^{\rho\sigma},
\label{SgH}
\end{align}
where the trace ${\rm Tr}$ is taken over all the particle representations. As usual, the Moyal-Weyl
$\star$-product is given by
$f\star g=f\left(\exp{\frac{i}{2}\,h\theta^{\m\n}\overleftarrow{\partial}_\m\overrightarrow{\partial}_\n}\right)g$, 
which implements the space-time noncommutativity as $x^\m\star x^\n-x^\n\star x^\m=ih\theta^{\m\n}$. 
The noncommutative deformation parameter $h=1/\Lambda^2_{\rm NC}$ sets the noncommutative scale. 
The noncommutative field strength $\hat F^{\m\n}$ depends on 
the enveloping-algebra valued noncommutative gauge field $\hat V_\m$ as 
\begin{align}
{\widehat F}_{\mu\nu}(x)
= \partial_{\mu}{\widehat V}_{\nu}
-\partial_{\nu}{\widehat V}_{\mu}-i[{\widehat V}_{\mu}\stackrel{\star}{,}{\widehat V}_{\nu}],
\label{F}
\end{align}
and in turn, $\hat V_\m$ depends on the ordinary gauge bosons through the Seiberg-Witten map
\begin{eqnarray}
\widehat V_\mu(x) &=&V_\mu(x) -\frac{h}{4} \theta ^{\sigma\rho}\left\{
V_\sigma(x), \partial _\rho V_\mu(x) +F_{\rho \mu}(x)\right\} + O(h^2)\,,
\label{SW}\\
V_\mu(x)&=&g'{\cal A}_\m(x) Y+g\sum_{a=1}^3 B_{\m,a}(x) T^a_L+g_s\sum_{b=1}^8G_{\m,b}(x)T^b_S\,,
\label{V}
\end{eqnarray}
where $V_\m(x)$
is the Standard Model gauge potential taking values in the Lie algebra of 
$SU(3)_C\times SU(2)_L\times U(1)_Y$. 
The pure gauge action defined by eq.~\eqref{SgH}, after expanding to order $h$ with 
the SW map of eq.~\eqref{SW} leads to
\begin{eqnarray}
S_g &=& {\rm Tr}\int \mathrm{d}^4x\,\left(-\frac{1}{2}
F_{\mu\nu}F^{\mu\nu}
+h\,\theta^{\mu\nu}\, (
\frac{a}{4}\, F_{\mu\nu}F_{\rho\sigma}
-F_{\mu\rho}F_{\nu\sigma} )F^{\rho\sigma}\right)\,.
\label{SgC}
\end{eqnarray} 
After taking traces in the above action over all massive particle representations with different 
quantum numbers that appear in the total lagrangian of the model with covariant derivatives acting on them,
the gauge action (\ref{SgC}) produces triple neutral 
gauge boson interactions \cite{Behr:2002wx,Melic:2005fm} which are not 
present in the mNCSM; 
in this paper we are interested in the $\ga\ga\ga$ and $Z\ga\ga$ couplings, 
which arise from the following terms in the lagrangian \cite{Buric:2007qx}
\begin{eqnarray}
{\cal L}^{\rm nmNCSM}_{\gamma\gamma\gamma}(a) &=& 
\frac{e}{4} \sin{2\theta_W} K_{\gamma\gamma\gamma} 
h\theta^{\rho\sigma} A^{\mu\nu} \left (a A_{\mu\nu} A_{\rho\sigma} 
- 4 A_{\mu\rho} A_{\nu\sigma} \right ),
\nonumber\\
{\cal L}^{\rm nmNCSM}_{Z\gamma\gamma}(a) &=&
\frac{e}{4} \sin{2\theta_W} K_{Z\gamma\gamma}  
h\theta^{\rho\sigma} \Big[ 2 Z^{\mu\nu} ( 2 A_{\mu\rho} A_{\nu \sigma} 
- aA_{\mu\nu} A_{\rho\sigma} ) 
+ 8 Z_{\mu\rho} A^{\mu\nu} A_{\nu\sigma} - 
a Z_{\rho\sigma} A_{\mu\nu}A^{\mu\nu} \Big ].
\label{Sgtheta}
\end{eqnarray}
See fig.~\ref{f:0} for details on 
the allowed values for the constants $K_{\ga\ga\ga}$ and $K_{Z\ga\ga}$ 
and their dependence on other parameters of the model \cite{Behr:2002wx}. The remarkable result of 
refs.~\cite{Buric:2006wm} is that the gauge action ~\eqref{SgC} is 
one-loop renormalisable up to order $h$ for $a=3$, with the $O(h)$ quantum 
corrections being finite. Thus we will be mainly interested in the $a=3$ case.
\section{Matter sector}

\subsection{Minimal fermion action}

Next we turn our attention to the fermion sector. In both the mNCSM and nmNCSM models, 
the chosen minimal fermion action was
\begin{eqnarray}
S_\psi^{min}=\idx\overline{\widehat\psi}\star i\hat\Dirac\star\widehat\psi+\hat S_{\rm Yukawa},
\quad \hat D_\m=\partial_\m-i \hat V_\m\star,
\label{Sfermin}
\end{eqnarray}
which depends on the ordinary fields $V_\m$ and $\psi$ through the SW map of 
eq.~\eqref{SW} and the fermion SW map
\begin{eqnarray}
{\widehat \psi}= \psi - 
\frac{h}{2}\,\theta^{\sigma\rho} \left(V_{\sigma}\partial_{\rho}
-\frac{i}{4}\,[V_{\sigma} , V_{\rho}] \right)\psi+O(h^2)\,.
\label{fermionSW}
\end{eqnarray}
The Yukawa sector in the NCSM, when expanded in terms of ordinary fields, 
involves quite complicated interactions \cite{Melic:2005fm}. However, if we consider the noncommutative 
contributions to the fermion field interactions in the QED sector only, 
(as turns out to be sufficient for our purpose of obtaining rate for
the quarkonia decay into two photons), 
we can use, as noted in \cite{Melic:2005fm} 
and used in \cite{Melic:2005hb}, the simplified fermion Lagrangian
\begin{align}
\nonumber
S^{min}_{\psi,A}&\sim\idx\,\overline{\widehat \psi}\star (i\hat\Dirac^A-m_f)\star\widehat \psi
= S_{\psi,A}^{SM}+S_{\psi,A}^\th+O(h^2),\\
 S_{\psi,A}^{SM}&=\int d^4x
 \overline{\psi} \, (i \slash{\!\!\!\!D}^A -m_f)\, \psi,\quad S_{\psi,A}^\th=-\frac{eh}{4}\int d^4x
 \, \overline{\psi}\, A_{\mu \nu} \,
 (i \theta^{\mu \nu \rho} \,  D_{\rho}^A -m_f \, \theta^{\mu \nu})\psi ,
\label{Sfsimp}
\end{align}
where $D_\m^A=\partial_\m-ie A_\m$ and 
\begin{equation}
 \theta^{\mu \nu \rho}=\theta^{\mu\nu} \gamma^{\rho}
+ \theta^{\nu \rho} \gamma^{\mu}
+ \theta^{\rho \mu} \gamma^{\nu}\,.
\label{thetamnr}
\end{equation}

\subsection{Deformed fermion action}

Motivated by the unknown renormalisability properties of the NC fermion sector, 
and since in the case of the gauge sector the addition of the higher order term dependent 
on the free deformation parameter 
$a$ made 
the model surprisingly well-behaved, we will consider adding deformation terms to 
the minimal fermion lagrangian (\ref{Sfsimp}), 
either to treat it in an effective theory approach or to prepare the grounds for future investigations 
of renormalisability. Thus, we set to find all the possible deformations 
contributing to the fermion action, and satisfying the following conditions:\\ 
(a) They are real and include two fermions fields, \\
(b) are invariant under 
noncommutative gauge transformations and thus involve star products, noncommutative gauge covariant 
derivatives and noncommutative field strengths,\\ 
(c) involve a contraction with a $\theta^{\m\n}$ tensor outside 
the star product ---as in the $a$-dependent term in eq.~\eqref{SgH}---,\\ 
(d) they do not alter the tree-level 2-point function of fermion propagator,\\ 
(e) they only generate P-symmetry violating contributions coming from the ordinary weak SU(2) gauge fields,\\
(f) include zero or positive powers of some mass parameter.

 In order to write down all the possible terms satisfying the conditions above, we have to consider 
 a basis of matrices in spinor space. Since the $\ga$ matrices are constrained by the identity 
 $\{\ga^\m,\ga^\n\}=2g^{\m\n}$, we can consider a basis formed by antisymmetrised products of $\ga$ matrices:
 \{$I,\ga^\m,\sigma^{\m\n},\ga^{\m\n\r},\ga^5$\}, where we are using the following definitions:
\begin{align}
 &\sigma^{\m\n}=\frac{1}{2}\big(\ga^\m\ga^\n-\ga^\n\ga^\m\big),
 \nonumber\\
&\ga^{\m\n\r}=\frac{1}{6}\big(\ga^\m\ga^\n\ga^\r+\ga^\r\ga^\m\ga^\n
+\ga^\n\ga^\r\ga^\m-\ga^\n\ga^\m\ga^\r-\ga^\r\ga^\n\ga^\m-\ga^\m\ga^\r\ga^\n\big),
\nonumber\\
&\ga^\m\ga^\n\ga^\r=g^{\m\n}\ga^\r-g^{\m\r}\ga^\n+g^{\n\r}\ga^\m-i\epsilon^{\m\n\r\s}\ga^5\ga_\s\,,
\nonumber\\
&\ga^5=-\frac{i}{4!}\epsilon_{\m\n\r\s}\ga^\m\ga^\n\ga^\r\ga^\s.
\label{gamma}
\end{align}

With this in mind, it can be seen that the terms satisfying the conditions stated above 
are given by sums of integrals of the following monomials $t_i$ multiplied by real coefficients:
\begin{align}
 \nonumber t_1&=h\theta^{\alpha\beta}\bar{\widehat\psi}\star\gamma^\mu 
 (\hat{\cal { D}}_\mu {\hat F}_{\alpha\beta})
\star {\widehat\psi}, 
& t_2&=h\theta^{\alpha\beta}\bar{\widehat\psi}\star\gamma_\beta(\hat{\cal { D}}^\mu {\hat F}_{\mu\a}) 
\star{\widehat\psi}, \\
 \nonumber t_3&=
 ih\theta^{\a\b}\bar{\widehat\psi}\star\ga^\r(2\hat F_{\a\b}\star \hat D_\r
 +(\hat{\cal { D}}_\r \hat F_{\a\b}))\star\widehat\psi, 
 & t_4&=ih\theta^{\a\b}\bar{\widehat\psi}\star\ga^\r(2\hat F_{\b\r}\star \hat D_\a
 +(\hat{\cal { D}}_\a \hat F_{\b\r}))\star\widehat\psi,\\
\nonumber t_5&=ih\theta^{\a\b}\bar{\widehat\psi}\star\ga_\b(2\hat F_{\m\a}\star \hat D^\m
+(\hat{\cal { D}}^\m \hat F_{\m\a}))\star\widehat\psi, 
& t_6&=i h\theta^{\alpha\beta}\bar{\widehat\psi}
\star{\ga_{\a\b}}^{\,\r}(\hat{\cal D}^\m \hat F_{\m\r})\star\widehat\psi,\\
\label{fH} t_7&=i h\theta^{\alpha\beta}\bar{\widehat\psi}\star{\ga_{\b}}^{\,\m\n}
(\hat{\cal D}_\a \hat F_{\m\n})\star\widehat\psi,
& t_8&=h\theta^{\a\b}\bar{\widehat\psi}\star{\ga_{\a\b}}^{\,\r}(2\hat F_{\m\r}\star \hat D^\m 
+(\hat{\cal { D}}^\m \hat F_{\m\r}))\star\widehat\psi, \\ 
\nonumber t_9&=h\theta^{\a\b}\bar{\widehat\psi}\star{\ga_{\b}}^{\,\m\n}
(2\hat F_{\m\n}\star \hat D_\a+(\hat{\cal { D}}_\a \hat F_{\m\n}))\star\widehat\psi, 
& t_{10}&=h\theta^{\a\b}\bar{\widehat\psi}\star{\ga_{\b}}^{\,\m\n}(2\hat F_{\n\a}\star \hat D_\m
+(\hat{\cal { D}}_\m \hat F_{\n\a}))\star\widehat\psi,\\
\nonumber t_{11}&=m\,h\,\theta^{\a\b}\bar{\widehat\psi}\star \hat F_{\a\b}\star\widehat\psi, 
& t_{12}&=im\,h\,\theta^{\a\b}\bar{\widehat\psi}\star{\sigma_{\beta}}^\r \hat F_{\a\r}\star\widehat\psi,\\
\nonumber t_{13}&=im\,h\,\tilde\theta^{\a\b}\bar{\widehat\psi}\star\ga^5 \hat F_{\a\b}\star\widehat\psi,
\end{align} 
with $\tilde\theta^{\a\b}\equiv\frac{1}{2}\epsilon^{\a\b\r\s}\theta_{\r\s}$,
and $ {\hat{D}}_\mu=\partial_\mu-i\hat V_\mu\star\,,\;\;
{\hat{\cal D}}_\mu=\partial_\mu-i[\hat V_\mu\stackrel{\star}{,}~~~]\,$.
Note that we can identify the mass parameter $m$ with any of the fermion masses $m_f$; 
this means no loss of generalisation because in principle we will be adding these 
mass-dependent terms to the action multiplied by arbitrary dimensionless coefficients. 

In order to check the C,P,T transformations of the above terms, we consider the following:
\begin{itemize}
 \item $\theta^{\m\n}$ transforms under discrete space-time symmetries as a U(1) 
 field strength $F^{\m\n}$  \cite{Aschieri:2002mc}.
\item The ordinary fields are the ones that define the theory, so that we have to deal 
with their C,P,T transformation properties. However, the analysis can be simplified because of  
the fact that the SW maps that we use are such that the C,P,T transformations of 
the noncommutative fields are equal to their commutative counterparts~\cite{Aschieri:2002mc}. 
However, there is a subtlety  since the noncommutative vector field in the matter representations, 
given the expansion of eq.~\eqref{V}, includes some chiral projectors that come together with 
the weak gauge fields $B_{\m,a}(x)$.
\end{itemize}
Taking this into account, it can be seen that the above terms are P invariant save for 
the contributions involving the ordinary SU(2) gauge fields, as was required from the start. 
Moreover, all terms in eqs.  (\ref{fH}) are CPT invariant except for $t_{11},t_{12},t_{13}$. 
In the terms $t_{11},t_{12},t_{13}$ the CPT violating 
contributions come exclusively from the chiral projectors of the weak gauge fields. 
However, in the strong and hypercharge sectors
the contributions from all terms in (\ref{fH}) remain CPT invariant.
Interestingly, the terms $t_1,t_2,t_8,t_9,t_{10},t_{12}$ originate both C and T violations in 
the strong and hypercharge sectors, with CT conserved in these sectors. 
All other violations of C, P or T come from the weak fields exclusively.
We will not worry  about these violations of discrete symmetries and will proceed considering 
all of the  terms in eq.~\eqref{fH}. Thus, we will consider an additional piece of 
the action given by a sum of the above terms,
\begin{equation}
 S^{H}_\psi(x_i)=\idx\sum_i x_i t_i,\quad x_i\in \mathbb{R}\,,
\label{Sx}
\end{equation}
where $x_i$'s are the fermion sector free deformation parameters, 
to be constrained via considerations of renormalisability
or by phenomenology.

Before plunging into the computation of the Feynman rules and their phenomenological application,
we would like to analyse 
whether any of the terms in eq.~\eqref{fH} can be reabsorbed by means of field redefinitions. 
This is interesting for two reasons:\\
- Possible future investigations about renormalisability would require to take into account 
 the effects of field redefinitions in the action, which we will compute here in the fermion sector.\\ 
- Field redefinitions should not affect S-matrix elements, and since we aim 
 to calculate quarkonia decay rates it is desirable to eliminate as many of the $x_i$ parameters 
 in eq.~\eqref{Sx} as possible. 

 To proceed, we consider the most general field redefinitions of order $\theta$ of 
 the commutative gauge and fermion fields, involving zero or positive powers of 
 the mass parameter and not affecting the tree-level propagator, 
 and compute the associated change in the action.
We parameterise the most general field redefinitions up to order $\theta$ as
\begin{equation}
 \delta v_\m=\sum y_i \delta^i{v}_\m,\quad \delta \psi=\sum z_i \delta^i{\psi},
 \label{fr1}
\end{equation}
where $y_i$ are real, $z_i$ complex, and $\delta^{i}v_\m$ and $\delta^{i}\psi$ are given next
\begin{eqnarray}
 \delta^1{v}_\m&=&\theta^{\a\b}{\cal D}_\m F_{\a\b},\;\;\; 
 \delta^2{v}_\m={\theta_\m}^{\,\a}{\cal D}^\n F_{\n\a}, 
 \nonumber\\ 
\delta^1{\psi}&=&\theta^{\a\b}F_{\a\b}\psi,\;\;\;\;\;\;\, 
\delta^2{\psi}=\theta^{\a\b}{\sigma_{\beta}}^\r F_{\a\r}\psi, 
\;\;\;\;\;
\delta^3{\psi}=\tilde\theta^{\a\b}\ga^5F_{\a\b}\psi\,.
\label{fr2}
\end{eqnarray}
Since the field redefinitions are of order $h$, to see their effect on the action 
to this same order we only need to compute the variation of the $O(h^0)$ action, 
which for fermions and vector fields is just 
\begin{eqnarray}
S_{v,\psi}^{SM}=S^{SM}_{g}+\int d^4x
 \overline{\psi} \, (i \slash{\!\!\!\!D} -m_f)\, \psi+S_{\rm Yukawa}\,; 
\label{SM}
\end{eqnarray} 
we will ignore 
 the Yukawa couplings since they do not contribute to the tree-level quarkonia 
 decay amplitudes that we want to compute. The change of the total action turns out to be
\begin{equation}
\delta\left(S^{SM}_{g}+\int d^4x
 \overline{\psi} \, (i \slash{\!\!\!\!D} -m_f)\, \psi\right)=\idx\sum_i (y_i\delta^{y_i}{\cal L}
 +\text{Re} z_i\,\delta^{z_i,R}{\cal L}+\text{Im}z_i\,\delta^{z_i,I}{\cal L})+O(h^2),
 \label{fr3}
\end{equation}
with
\begin{align}
 \delta^{y_1}{\cal L}&=t_1, & \delta^{y_2}{\cal L}&=-t_2, & \delta^{z_1,R}{\cal L}&=t_3-2t_{11},
 \nonumber\\
 \delta^{z_1,I}{\cal L}&=-t_1, &\delta^{z_2,R}{\cal L}
 &=-t_4+t_5-\frac{1}{2}t_7, & \delta^{z_2,I}{\cal L}
 &=-\frac{1}{2}t_1-t_2-t_{10}-2t_{12},
 \nonumber\\
\delta^{z_3,R}{\cal L}&=-t_8-t_9, & \delta^{z_3,I}{\cal L}
&=-t_6-t_7-2t_{13},
 \label{fr4}
\end{align}
where the $t_i$'s are given in eq.~\eqref{fH}. In the language of 
eq.~\eqref{Sx}, the  above result is equivalent to the following change in the action $S_\psi^{H}(x_i)$:
\begin{align}
 S_\psi^{H}(x_i)\longrightarrow S_\psi^{H}(x_i+\delta x_i)\,,
  \label{fr5}
\end{align}
with
\begin{align}
 \delta x_1&=y_1-\text{Im}z_1-\frac{1}{2}\text{Im}z_2, 
 & \delta x_2&=-y_2-\text{Im}z_2, & \delta x_3&=\text{Re}z_1\,,
 \nonumber\\
\delta x_4&=-\text{Re}z_2, & \delta x_5&=\text{Re}z_2\,, 
&\delta x_6&=-\text{Im}z_3\,,
\nonumber\\
\delta x_7&=-\frac{1}{2}\text{Re}z_2-\text{Im}z_3, 
& \delta x_8&=-\text{Re}z_3, & \delta x_9&=-\text{Re}z_3\,,
\nonumber\\
\delta x_{10}&=-\text{Im}z_2, & \delta x_{11}&=-2\text{Re}z_1, 
& \delta x_{12}&=-2\text{Im}z_2\,,
\nonumber\\
\delta x_{13}&=-2\text{Im}z_3.
 \label{fr6}
\end{align}

In order to check whether the field redefinitions above can be used 
to eliminate some of the terms $t_i$ of the action in eq.~\eqref{Sx}, 
we have to examine the system of equations that follows
\begin{align}
 \delta x_i[y_j,\text{Re}z_j,\text{Im}z_j]=-x_i\,,
  \label{fr7}
\end{align}
where $x_j$ are to be treated as fixed and $y_i,\text{Re}z_i,\text{Im}z_i$ as the unknown variables. 
Now, the $13\times 8$ matrix associated to the previous linear system of 
equations can be seen to be of rank 7. This means that at most 7 of the $t_i$ 
terms in the action can be eliminated with adequate field redefinitions, with 6 terms 
remaining. It is easily seen that a  viable choice of 6 terms that survive 
the field redefinitions is given by $t_4,t_5,t_6,t_8,t_{10},t_{11}$ --note that, as was commented before, 
$t_{11}$ violates CPT in the weak sector and $t_8,t_{10}$ violate C and T in the strong and hypercharge sectors.

\section{Feynman rules to compute S-Matrix elements}

From the discussions above it follows that, in order to compute the amplitudes of the desired physical processes, 
the relevant pieces of 
the action are those given by the ordinary Standard Model fermion-photon 
and fermion-Z boson interactions, plus the following noncommutative 
$\th$-dependent interactions: the three boson interactions of eq.~\eqref{Sgtheta}, 
the fermion-photon interaction in eq.~\eqref{Sfsimp}, and the contribution 
to the action of the terms of eq.~\eqref{Sx} surviving the fermion field redefinitions, i.e., 
\begin{equation}
S_\psi^{H}(x_i)=\idx \Big[x_4 t_4+x_5 t_5+x_6 t_6+x_8 t_8+x_{10} t_{10}+x_{11}t_{11}\Big].
\label{Stx}
\end{equation}
 The noncommutative interactions originate, among other vertices, 
 a triple neutral gauge boson vertex, a 2 fermion-photon vertex and a 2 fermion-2 boson vertex. 
From the modified gauge and fermion actions, (\ref{SgH},\ref{SgC},\ref{Sgtheta},\ref{Sfsimp},\ref{Stx}),
\begin{equation}
 S=S_g+S_{\psi}=S_g^{min}+S_g^H(a)+ S_\psi^{min} + S_\psi^{H}(x_i)+O(h^2),
\label{mfS}
\end{equation}
we obtain the following Feynman rules:\\

\begin{minipage}{0.18\textwidth}\vskip0.2cm
\includegraphics[scale=0.43]{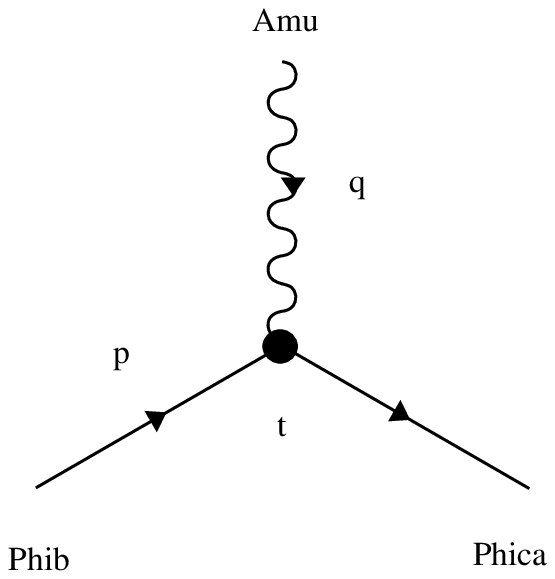}
\end{minipage}%
\begin{minipage}{0.795\textwidth}
 \begin{align}
  \leftrightarrow  &\, ie \, e_q\,
  \Big[\gamma^{\mu}-i
 \frac{h}{2} \, k_{\nu}
   \big( \theta^{\mu \nu \rho}_H
 \, p_{\rho}
\, +\,\tilde\theta^{\mu \nu \rho}_H\,k_\rho
-\,(1+4x_{11})\, m_f\,\theta^{\mu \nu}\,\big)
  \Big]_{\a\b}\,,
  \nonumber\\
&\theta^{\mu \nu \rho}_H=
  \theta^{\mu \nu\rho}
-4x_4(\th^{\r\m}\ga^\n+\theta^{\nu \rho} \gamma^{\mu})-4
x_5(-g^{\n\r}\th^{\m\a}\ga_\a+ g^{\mu\rho}\theta^{\nu\alpha}\gamma_\alpha)
\nonumber\\
&\phantom{\theta^{\mu \nu \rho}_H=}+8 i x_8 g^{\m\r}\theta^{\s\eta}{\ga_{\s\eta}}^{\,\n}
+8i x_{10}\theta^{\n\s}{\ga_\s}^{\,\r\m},
\nonumber\\
&\tilde\theta^{\mu \nu \rho}_H=
  \frac{1}{2}\big(\theta^{\mu \nu\rho}_{H}-\theta^{\mu \nu \rho}\big)
  -4x_6g^{\r\m}\th^{\s\eta}{\ga_{\s\eta}}^{\,\n},\;\;\;\;\;
  e_q=\frac{2}{3},-\frac{1}{3},\,\,\,{\rm for \;\; quarks},
\label{Fr1}
 \end{align}
\end{minipage}
\vspace{.8cm}

\begin{minipage}{0.21\textwidth}
\includegraphics[scale=0.43]{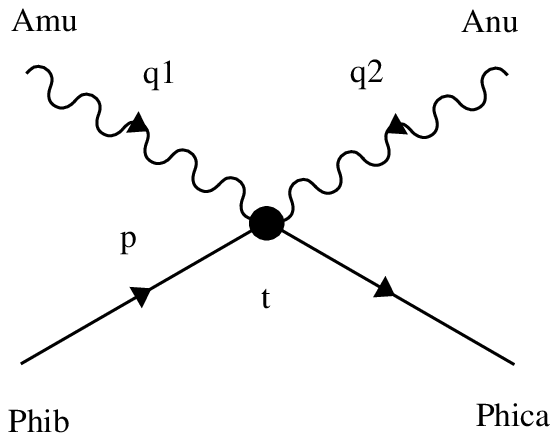}
\end{minipage}
\begin{minipage}{0.755\textwidth}
\begin{align}
 \leftrightarrow-\frac{h}{2}\, e^2 \, e_q^2\,
[\theta^{\mu \nu \rho}_H]_{\a\b}\,(k_1-k_2)^{\rho}\,,
\label{Fr2}
\end{align}
\end{minipage}
\vspace{.8cm}

\begin{minipage}{0.2\textwidth}
\includegraphics[scale=0.43]{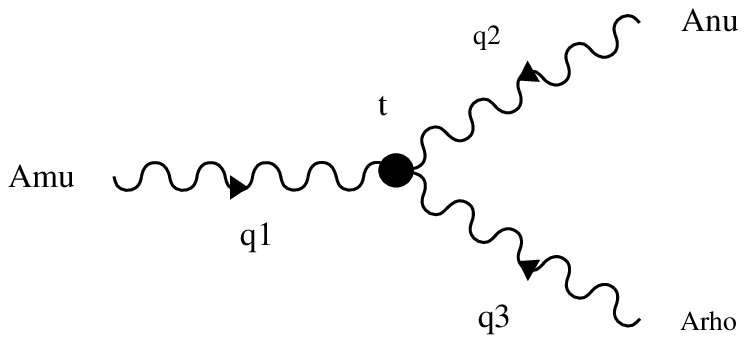}
\end{minipage}
\begin{minipage}{0.775\textwidth}
\begin{align}
 \leftrightarrow 2e\sin(2\th_W)K_{\ga\ga\ga}\,h\,\Theta_3^{\mu\nu\rho}[a;k_1,k_2,k_3]\,,
 \label{Fr3}
\end{align}
\end{minipage}
\vspace{.8cm}

\begin{minipage}{0.2\textwidth}
\includegraphics[scale=0.43]{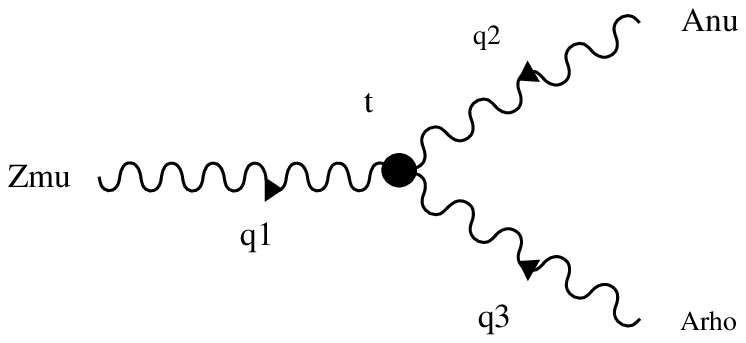}
\end{minipage}%
\begin{minipage}{0.78\textwidth}
\begin{align}
 \leftrightarrow -2e\sin(2\th_W)K_{Z\ga\ga}\,h\,\Theta_3^{\mu\nu\rho}[a;k_1,k_2,k_3]\,.
 \label{Fr4}
\end{align}
\end{minipage}

\vskip 1cm

Here, $\th^{\m\n\r}$ was given in eq.~\eqref{thetamnr}, while the tensor
$\Theta_3^{\mu\nu\rho}[a;k_1,k_2,k_3]$, is given by \cite{Buric:2007qx}:
\begin{eqnarray}
{\Theta^{\mu\nu\rho}_3}[a;k_1,k_2,k_3]&=&
-\,(k_1 \theta k_2)\,
[(k_1-k_2)^\rho g^{\mu \nu} +(k_2-k_3)^\mu g^{\nu \rho} + (k_3-k_1)^\nu g^{\rho \mu}]
\label{3g} \\
& -&
\,\theta^{\mu \nu}\,
[ k_1^\rho \, (k_2 k_3) - k_2^\rho \, (k_1 k_3) ]
-\,\theta^{\nu \rho}\,
[ k_2^\mu \, (k_3 k_1) - k_3^\mu \, (k_2 k_1) ]
-\,\theta^{\rho \mu}\,
[ k_3^\nu \, (k_1 k_2) - k_1^\nu \, (k_3 k_2) ]
\nonumber \\
& +&
\,(\theta k_2)^\mu \,\left[g^{\nu \rho}\, k_3^2 - k_3^\nu k_3^\rho\right]
+(\theta k_3)^\mu\,\left[g^{\nu \rho}\, k_2^2 - k_2^\nu k_2^\rho\right]
\nonumber \\
& +&
\,(\theta k_3)^\nu \,\left[g^{\mu \rho}\, k_1^2 - k_1^\mu k_1^\rho \right]
+(\theta k_1)^\nu \,\left[g^{\mu \rho}\, k_3^2 - k_3^\mu k_3^\rho \right]
\nonumber \\
& +& 
\,(\theta k_1)^\rho \,\left[g^{\mu \nu}\, k_2^2 - k_2^\mu k_2^\nu \right]
+(\theta k_2)^\rho \,\left[g^{\mu \nu}\, k_1^2 - k_1^\mu k_1^\nu \right]
\nonumber \\
&+&
(a-1)\Big((\theta k_1)^{\mu}\,\left[g^{\nu \rho}\,(k_3 k_2)-k_3^\nu k_2^\rho\right]
+(\theta k_2)^{\nu}  \left[g^{\mu \rho}(k_3 k_1)-k_3^\mu k_1^\rho\right]
+(\theta k_3)^{\rho} \left[g^{\mu \nu}(k_2 k_1)-k_2^\mu k_1^\nu\right]\Big)\,.
\nonumber
\end{eqnarray}

\section{Application of the proposed framework to quarkonia decays: $\overline{q}q_{1} \rightarrow \gamma\gamma$}

\subsection{Quarkonia decay amplitudes}

\begin{figure}[h]
\begin{center}
\includegraphics[scale=.9]{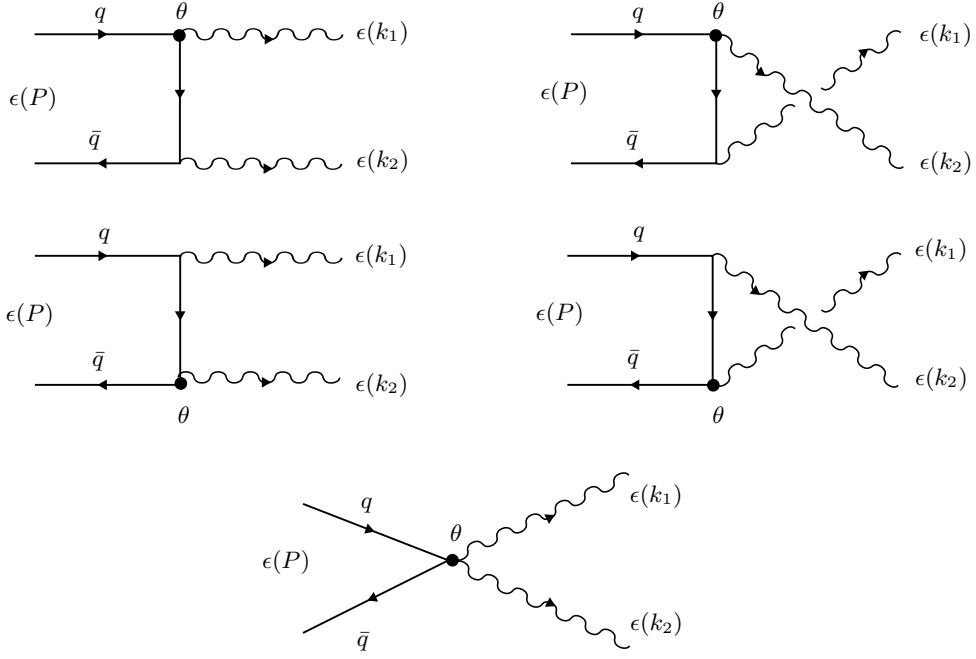}
\end{center}
\caption{Contributions to the 
${\cal A}_1(\overline{q}q_{1} \rightarrow \gamma \gamma)$ amplitude.}
\label{f:1}
\end{figure}

\begin{figure}
\begin{center}
\hspace{-1cm}
\includegraphics[scale=.6]{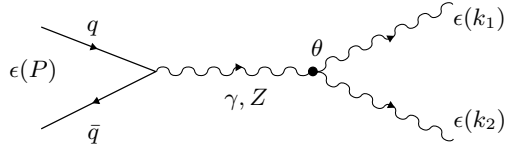}
\end{center}
\caption{Additional contributions to the 
${\cal A}_2(\overline{q}q_{1} \rightarrow \gamma \gamma)$ amplitude.}
\label{f:2}
\end{figure}

The diagrams that contribute to the quarkonia decay amplitude 
are shown in figures \ref{f:1} and \ref{f:2}.
Note that the vertices with fermions including noncommutative effects (with black dots), only involve photons; 
 that's why we may safely use \eqref{Sfsimp}. The diagram  in fig.~\ref{f:2} 
 also involves a fermion-fermion-Z boson vertex, 
 but it is given by the ordinary SM contribution.

Using the Feynman rules
(\ref{Fr1}-\ref{Fr4}) 
obtained from the modified gauge (\ref{Sgtheta}) and fermion (\ref{mfS}) actions, we have evaluated
the diagrams from Figs. \ref{f:1} and \ref{f:2}, yielding amplitudes that we call
 ${\cal A}_1$ and ${\cal A}_2$, respectively.

As in ref.~\cite{Melic:2005hb}, in order to hadronise 
the free quarks into the quarkonium bound state, 
we apply
 the following prescription for the transition amplitude of the operator 
 $q^{\alpha}_i \overline{q}^{\beta}_j$ ($q =c,b$ and $i,j$ are colour indices) 
 from the vacuum to the quarkonium state:
\begin{eqnarray}
\langle 0 | q^{\alpha}_i \overline{q}^{\beta}_j | \overline{q}q_{1}(P) \rangle  =  
- \frac{|\Psi_{\overline{q}q_1}(0)|}{\sqrt{12 M}} 
\left[(\slash{\!\!\!\!P} + M) \slash{\!\!\!\epsilon}\right]^{\alpha\beta}
\, \delta_{ij}\,,
\label{vq}
\end{eqnarray}
where $|\Psi_{\overline{q}q}(0)|$ represents the quarkonia wave function at the origin 
defined in \cite{Melic:2005hb},
\begin{eqnarray}
|\Psi_{\overline{q}q_1}(0)|^2  = \frac{\Gamma(\overline{q}q_{1} \to \ell^+ \ell^-) 
M^2}{ 16 \pi \alpha^2 e_q^2 }\,.
\label{wf}
\end{eqnarray} 

We use a collinear approximation for the quarks in the quarkonium state, 
and we identify the mass of the quarkonium as $M=2m_q$. The resulting amplitudes are shown next:
\begin{eqnarray}
{\cal A}_1(x_4,x_5,x_{11}) &=&
ih\pi{4\sqrt{3M}}\alpha e_q^2 |\Psi_{\overline{q}q_1}(0)| 
\epsilon_{\mu}(k_1) \epsilon_{\nu}(k_2) \epsilon_{\rho}(P)
\label{A1}
\Big \{
(k_2 - k_1)^{\rho} \left[(1-4x_4)\left( \theta^{\mu\nu} - 2g^{\mu\nu} 
\frac{(k_1 \theta k_2) }{M^2} \right )\right.\nonumber\\
&-&\left.4x_5\left( \theta^{\mu\nu} - 
\frac{k_2^{\mu}(k_1 \theta)^{\nu}-k_1^{\nu}(k_2 \theta)^{\mu}}{M^2}\right) \right] 
 -   4x_5(P\theta)^{\rho}\left(g^{\mu\nu}-\frac{2}{M^2}k_2^{\mu} k_1^{\nu}\right)  
\nonumber \\
& + &  (1-4x_4-2x_5) \left[2 g^{\mu \rho} \left( (k_1 \theta)^{\nu} 
- 2 k_1^{\nu} \frac{(k_1 \theta k_2) }{M^2} \right) 
 + 2 g^{\nu \rho} \left ( (k_2 \theta)^{\mu}  
+  2 k_2^{\mu} \frac{(k_1 \theta k_2) }{M^2} \right)\right]\\
\nonumber
&+& (4x_{11})\left[(k_1\theta)^{\mu}\left(g^{\nu\rho}+\frac{k_1^{\rho} k_1^{\nu}-k_2^{\rho} k_1^{\nu}}{M^2}\right)
+ (k_2\theta)^{\nu}\left(g^{\mu\rho}+\frac{k_2^{\rho} k_2^{\mu}-k_1^{\rho} k_2^{\mu}}{M^2}\right)\right]
\Big\} \,,
\nonumber 
\end{eqnarray}
which turns out to be independent of $x_6,x_8,x_{10}$,
due to the external momenta being on-shell, and

\begin{eqnarray}
{\cal A}_2(a) &=& -ih\pi\frac{16\sqrt{3}}{M^{3/2}}\alpha  |\Psi_{\overline{q}q_1}(0)| 
\,\epsilon_{\mu}(k_1) \epsilon_{\nu}(k_2) \epsilon_{\rho}(P)
\left [ e_q\,\sin2\theta_{W} K_{\gamma\gamma\gamma} + 
\left ( \frac{M}{M_Z} \right )^2 c_V^q K_{Z\gamma\gamma} \right ]
\Theta^{\mu\nu\rho}_3[a;P,-k_1,-k_2]
\nonumber \\
&=& -ih\,\pi\,{8\sqrt{3M}}\,\alpha \, |\Psi_{\overline{q}q_1}(0)| 
\,\epsilon_{\mu}(k_1) \epsilon_{\nu}(k_2) \epsilon_{\rho}(P)
\left [ e_q\,\sin2\theta_{W} K_{\gamma\gamma\gamma} + 
\left ( \frac{M}{M_Z} \right )^2 c_V^q K_{Z\gamma\gamma} \right ]
\nonumber \\ 
& & \times \Big \{
(k_2 - k_1)^{\rho} \left ( \theta^{\mu\nu} - 2g^{\mu\nu} 
\frac{(k_1 \theta k_2) }{M^2} \right )   
\nonumber \\
& &  + 2 g^{\mu \rho} \left ( (k_1 \theta)^{\nu} 
- 2 k_1^{\nu} \frac{(k_1 \theta k_2) }{M^2} \right ) 
+ 2 g^{\nu \rho} \left ( (k_2 \theta)^{\mu}  
+  2 k_2^{\mu} \frac{(k_1 \theta k_2) }{M^2} \right  )-
(a-1)\left[(P\theta)^{\rho}\left(g^{\mu\nu}-\frac{2}{M^2}k_2^{\mu} k_1^{\nu}\right)\right.
\nonumber \\
& &  \left.
+(k_1\theta)^{\mu}\left(g^{\nu\rho}+\frac{k_1^{\rho} k_1^{\nu}-k_2^{\rho} k_1^{\nu}}{M^2}\right)
+(k_2\theta)^{\nu}\left(g^{\mu\rho}+\frac{k_2^{\rho} k_2^{\mu}-k_1^{\rho} k_2^{\mu}}{M^2}\right)\right]
\Big\} \, .
\label{AF2} 
\end{eqnarray}
The coupling constants appearing in the above amplitudes are evaluated 
at the $M_Z$ scale \cite{Behr:2002wx,Amsler:2008zz}. Here $P=k_1+k_2$,
$(k_i \theta)^{\mu} = k_{i\,\nu}\theta^{\nu\mu}$ and 
$k_1\theta k_2 = k_{1\,\mu} \theta^{\mu\nu} k_{2\,\nu}$, while
$M$ and $P$ are the mass and 
the total momentum of the discussed quarkonium state, respectively.

Both of the above amplitudes satisfy separately 
the usual Ward identities that follow from gauge invariance and the pole structure of the diagrams. 
Note that this would have been a problematic issue had we included in our deformed fermion lagrangian 
terms that contributed to the tree level fermion propagator.
It is also worth noticing that, despite the dependence of the Feynman rules of 
eqs.~\eqref{Fr1} and \eqref{Fr2} on the parameters $x_6,x_8,x_{10}$ --where $x_8$ and $x_{10}$ 
were associated to C, T violating contributions in the strong and hypercharge sectors 
--this dependence disappears in the amplitude ${\cal A}_1$ after evaluating 
the external momenta on-shell. In fact, the amputated Green function evaluated 
at generic momenta can be seen to depend on $x_6,x_8,x_{10}$,
thus not contradicting the expectations of ref. \cite{Grimstrup:2002af}, which stated the necessity of adding
a term involving $\gamma^{\mu\nu\rho}$ to get one-loop renormalisability. Indeed,
despite not contributing to the calculated amplitude, 
the associated terms in the basis of eq.~\eqref{fH} might influence other 
S-matrix elements or may be relevant to renormalisability properties. It is also worth to notice 
the dependence of the amplitude ${\cal A}_1$ in eq.~\eqref{A1} on the parameter $x_{11}$, 
which induces CPT violations in the weak sector. However, as can be seen from 
the diagrams of figure \ref{f:1}, 
the noncommutative vertices with fermion fields appearing in them only involve photons
---no weak fields---, 
so that our computation is not affect by CPT violating interactions. 

In ref.~\cite{Melic:2005hb}, when computing the amplitudes for 
the above diagrams in the nmNCSM without the 
$a$-deformation of eq.~\eqref{SgH} and without the additional deformations, $x_i$, of the fermionic 
terms,~\eqref{Stx}, it was found that the on-shell, amputated ${\cal A}_1$ amplitude, corresponding 
to the sum of the five  quark-exchange diagrams in Fig. \ref{f:1},   
was proportional to the on-shell, amputated amplitude ${\cal A}_2$ of the boson-exchange Green functions 
represented in fig. \ref{f:2}, which could hint at 
a possible symmetry. This was made manifest by the fact that 
the S-matrix amplitudes in ref.~\cite{Melic:2005hb} are given by some factor squared.  
In the present calculation, for $a=3$, which is 
the one making the gauge sector of deformed nmNCSM one-loop renormalisable, 
this result cannot be recovered for any value of the free deformation parameters 
$x_4,x_5,x_6,x_8,x_{10},x_{11}$; in fact proportionality is only achieved for $a=1$ and $x_i=0$.
The fact that we eliminated some of the 
freedom parameters $x_i$ should not be relevant to this result, 
since we are concerned with on-shell, amputated amplitudes which are not affected by 
the field redefinitions that allowed us to eliminate some of the $x_i$.

With respect to the possibility of considering terms deforming the tree-level fermion propagators, 
which have been ignored here, we would like to point out again 
that they would not only modify the on-shell conditions, 
but also cast serious doubts about the validity of the usual procedure of constructing 
the S-matrix amplitudes in terms of amputated diagrams; further considerations on 
these issues are out of the scope of this paper.

\subsection{Quarkonia decay rate}

To obtain the quarkonia decay rates we have to compute
\begin{eqnarray}
\Gamma(\overline{q}q_{1} \to \gamma \gamma )=
\frac{1}{2E_{\overline{q}q_{1}}}\frac{1}{4\pi^2}
\int \frac{d^3 k_1}{2E_1}\frac{d^3 k_2}{2E_2}\delta^4(P-k_1-k_2)
\frac{1}{2s_i+1}\sum_{spins}^{} {|{\cal M}|}^2\,\frac{1}{2}\,,
\label{rate}
\end{eqnarray}
where $\cal M$ is the total amplitude, i.e. ${\cal M}={\cal A}_1+{\cal A}_2$.
Summing over the initial spins and averaging over their final values in 
the square absolute value of the amplitude gives 

\begin{align}
\sum_{spins} {|{\cal M}|}^2=&\frac{3 e^4 M^3 |\Psi| ^2}{\left(M^2-M_Z^2\right){}^2}
 \Big[A(\theta_{\a\b}\theta^{\b\a})+B(\theta_{\a\r}{\theta^{\r}}_{\,\b} k_1^\a k_2^\b)
 +C(\theta_{\a\r}{\theta^{\r}}_{\,\b}k_1^\a k_1^\b+\theta_{\a\r}{\theta^{\r}}_{\,\b}k_2^\a k_2^\b)
 +D(\theta_{\a\b}k_1^\a k_2^\b)^2\Big];
\label{Asquare1}
\end{align}
\begin{align}
A&=-\Big[-2 C_v K_{\ga\ga Z} M^2
+e_q^2(4 x_4+4 x_5-1) \left(M^2-M_Z^2\right)+2e_q \sin \left(2 \theta _W\right) K_{\ga\ga\ga} 
\left(M^2-M_Z^2\right)\Big]{}^2,
\label{MsquareA}\\
B&=\frac{8}{M^2} \Big[
\left(5 a^2-22 a+25\right) C_v^2 K_{\ga\ga Z}^2 M^4
\nonumber\\
&+2 e_q^2 C_v K_{\ga\ga Z} \left(M^2-M_Z^2\right) M^2
\Big(-28 x_4-28 x_5+a (12 x_4+12 x_5-4 x_{11}-3)+12 x_{11}+7\Big)
\nonumber\\
 &+e_q^2 \sin ^2\left(2 \theta _W\right) K_{\ga\ga\ga}^2 \left(M^2-M_Z^2\right){}^2\left(5 a^2-22 a+25\right) 
 \nonumber\\
&+2 e_q^4 \Big(16 x_4^2+8 (4 x_5-2 x_{11}-1) x_4+18 x_5^2+2 x_{11}^2+4 x_{11}-4 x_5 (3 x_{11}+2)+1\Big) 
\left(M^2-M_Z^2\right){}^2
\nonumber\\
&-2 \sin \left(2 \theta _W\right) K_{\ga\ga\ga} \left(M^2-M_Z^2\right) 
\Big(e_q \left(5 a^2-22 a+25\right) C_v K_{\ga\ga Z} M^2
\nonumber\\
&+e_q^3 \Big(-28 x_4-28 x_5+a (12 x_4+12 x_5-4 x_{11}-3)+12 x_{11}+7\Big) \left(M^2-M_Z^2\right)\Big)
\Big]\,,
\label{MsquareB}
\end{align}
\begin{align}
C&=\frac{8}{M^2} \Big[\left(3 a^2-10 a+11\right) C_v^2 K_{\ga\ga Z}^2 M^4
\nonumber\\
&+2 e_q^2 C_v  K_{\ga\ga Z} \left(M^2-M_Z^2\right) M^2
\Big(-12 x_4-11 x_5+a (4 x_4+5 x_5-3 x_{11}-1)+5 x_{11}+3\Big)
\nonumber\\
&+e_q^2 \sin ^2\left(2 \theta _W\right) K_{\ga\ga\ga}^2 \left(M^2-M_Z^2\right){}^2\left(3 a^2-10 a+11\right)
\nonumber\\
& +e_q^4 \left(M^2-M_Z^2\right){}^2
\Big(16 x_4^2+8 (3 x_5-x_{11}-1) x_4+12 x_5^2+4 x_{11}^2+2 x_{11}-2 x_5 (4 x_{11}+3)+1\Big) 
\nonumber\\
&-2 \sin \left(2 \theta _W\right) K_{\ga\ga\ga} 
\left(M^2-M_Z^2\right) \Big(e_q\left(3 a^2-10 a+11\right) C_v K_{\ga\ga Z} M^2
\nonumber\\
&+e_q^3 \Big(-12 x_4-11 x_5+a (4 x_4+5 x_5-3 x_{11}-1)+5 x_{11}+3\Big) \left(M^2-M_Z^2\right)\Big)\Big]\,,
\label{MsquareC}\\
D&=-\frac{16}{M^4}\Big[\left(3 a^2-14 a+15\right) C_v^2 K_{\ga\ga Z}^2 M^4
\nonumber\\
&+4 e_q^2 C_v K_{\ga\ga Z} \left(M^2-M_Z^2\right) M^2\Big(-8x_4-8x_5+a(4x_4+4x_5-2x_{11}-1)+6 x_{11}+2\Big) 
\nonumber\\
&+e_q^2 \sin ^2\left(2 \theta _W\right) K_{\ga\ga\ga}^2 \left(M^2-M_Z^2\right){}^2\left(3 a^2-14 a+15\right)
\nonumber\\
 &+e_q^4 \left(M^2-M_Z^2\right){}^2 
 \Big(16 x_4^2+8 (4 x_5-4 x_{11}-1) x_4+20 x_5^2+4 x_{11}^2+8 x_{11}-8 x_5 (3 x_{11}+1)+1 \Big)
 \nonumber\\
 &-2 \sin \left(2 \theta _W\right) K_{\ga\ga\ga} \left(M^2-M_Z^2\right) 
 \Big(e_q (3 a^2-14 a+15) C_v K_{\ga\ga Z} M^2
\nonumber\\
&+2 e_q^3 \Big(-8 x_4-8 x_5+a (4 x_4+4 x_5-2 x_{11}-1)+6 x_{11}+2\Big) \left(M^2-M_Z^2\right)\Big)\Big]\,.
\label{MsquareD}
\end{align}
Considering the quarkonia at rest and using the phase space integrals
\begin{eqnarray}
(-\theta^2)\int \frac{d^3 k_1}{2E_1}\frac{d^3 k_2}{2E_2}\delta^4(P-k_1-k_2)
&=&\frac{2}{\Lambda^4_{\rm NC}}(\vec{B}_{\theta}^2 - \vec{E}_{\theta}^2)\frac{\pi}{2}\,,
\nonumber\\
(P\theta^2 P)\int \frac{d^3 k_1}{2E_1}\frac{d^3 k_2}{2E_2}\delta^4(P-k_1-k_2)
&=&\frac{M^2}{\Lambda^4_{\rm NC}}(\vec{E}_{\theta}^2)\frac{\pi}{2}\,,
\nonumber\\
\int \frac{d^3 k_1}{2E_1}\frac{d^3 k_2}{2E_2}\delta^4(P-k_1-k_2)(k_1\theta^2 k_1)
&=&\frac{M^2}{\Lambda^4_{\rm NC}}(\vec{E}_{\theta}^2 + \vec{B}_{\theta}^2)\frac{\pi}{12}\,,
\nonumber\\
\int \frac{d^3 k_1}{2E_1}\frac{d^3 k_2}{2E_2}\delta^4(P-k_1-k_2)(k_1\theta k_2)^2
&=&\frac{M^4}{\Lambda^4_{\rm NC}}( \vec{E}_{\theta}^2)\frac{\pi}{24}\,,
\label{psint}
\end{eqnarray}
starting from (\ref{rate}), for general $a, x_i$, we obtain the following decay rate 
\begin{eqnarray}
\Gamma^{\rm nmNCSM}(\overline{q}q_{1} \to \gamma \gamma )=
\frac{\pi}{24}\frac{\alpha^2\,M^2|\Psi|^2}{(M^2-M_Z^2)^2\,\Lambda^4_{\rm NC}}
\left[\Big(24A+4M^2(B+C)+M^4D\Big)\vec{E}_{\theta}^2 - \Big(24A+2M^2(B-2C)\Big)\vec{B}_{\theta}^2\right]\,.
\label{ratef}
\end{eqnarray}
The above rate turns into eq.(12) from \cite{Melic:2005hb} for $a=1,\, x_i=0$.  
We would like to analyse eq.~\eqref{ratef} and study the effects of having added 
to the action of the nmNCSM  
the extra terms that have been discussed in the previous sections, in order to compare 
the results with those calculated in ref.~\cite{Melic:2005hb}. In order to obtain some numerical values, 
we will look for the maxima and minima of the decay rates in the allowed region for 
$K_{Z\ga\ga}$ and $K_{\ga\ga\ga}$ -see figure \ref{f:0}- with the assumption that 
the $x_i$ parameters are ``natural'' 
and only take values between zero and one, 
and with two possible scenarios for the dimensionless constants $\vec{E}^2_\th$ and $\vec{B}^2_\th$: 
either both of them are of order one (space-time and space-space noncommutativity), 
or $\vec{E}^2_\th=0$ and $\vec{B}^2_\th$ is of order one (only space-space noncommutativity). 
Moreover, as was done in ref.~\cite{Melic:2005hb}, 
we will consider that the scale of noncommutativity varies between 
$\Lambda_{\rm NC}=0.25$ TeV and $\Lambda_{\rm NC}=1$ TeV.

In order to compute branching ratios, we use the following 
data taken from  \cite{Amsler:2008zz}: in the $J/\psi$ case,
$\Gamma^{\rm exp.}(J/\psi \to e^+ e^-)=(5.55 \pm 0.14 \pm 0.02)\, {\rm keV}$ and 
$\Gamma^{\rm exp.}_{\rm tot}(\Upsilon)=(93.2 \pm 2.1) \, {\rm keV}$, whereas 
for the $\Upsilon$ case, $\Gamma^{\rm exp.}(\Upsilon \to e^+ e^-)=(1.340\pm 0.018)\, {\rm keV}$ 
and $\Gamma^{\rm exp.}_{\rm tot}(\Upsilon)=(54.02 \pm 1.25) \, {\rm keV}$. 
Recall that the wavefunction at the origin $\Psi(0)$ is related 
to the lepton decay rate by eq.~\eqref{wf}.

First, since in the minimal NCSM (mNCSM) there are no $Z\ga\ga$ and $\ga\ga\ga$ couplings, 
the formula \eqref{ratef} can be used to recover the branching rations 
in the minimal NCSM --see ref.~\cite{Melic:2005hb}, eqs.~(19,20)-- by taking $K_{Z\ga\ga\ga}=K_{\ga\ga\ga}=0$, 
which yields, for $\vec{E}_\th^2=\vec{B}_\th^2=1$ and $0.25 \text{  TeV}\leq\Lambda_{\rm NC}\leq 1 \text{  TeV}$, 
\begin{equation}
 5.1\cdot 10^{-13}\lesssim BR_{[J/\psi\rightarrow\ga\ga]}^{mNCSM}\lesssim1.3\cdot 10^{-10}, 
 \quad 4.6\cdot 10^{-12}\lesssim BR_{[\Upsilon\rightarrow\ga\ga]}^{mNCSM}\lesssim1.2\cdot 10^{-9}.
\label{BRmNCSM}
\end{equation}
The values at  $\vec{E}_\th^2=0$, $\vec{B}_\th^2=1$  are suppressed by a factor of $3/10$.

In ref. \cite{Melic:2005hb}, the computation in undeformed nmNCSM was also done 
for $\vec{E}_\th^2=\vec{B}_\th^2=1$, yielding
\begin{eqnarray}
&&\frac{\Gamma^{a=1}(\overline{q}q_{1} \to \gamma \gamma )}
{\Gamma(\overline{q}q_{1} \to \ell^+ \ell^-)}
=
\frac{5}{24}\,e^2_q\,\left(\frac{M}{\Lambda_{\rm NC}} \right)^4
\left[1 -\frac{2}{e_q} \sin 2\theta_W K_{\gamma\gamma\gamma} 
- 
\frac{2}{e^2_q}\left(\frac{M}{M_Z} \right)^2 c_V^q K_{Z \gamma\gamma }\right]^2 \,. 
\label{pdwr}  
\end{eqnarray}
This corresponds to setting $a=1,\;x_i=0$ in the deformed version of the nmNCSM of this paper.

Computing the maximal values of the above rate for 
$0.25 \text{  TeV}\leq\Lambda_{\rm NC}\leq 1 \text{  TeV}$ yields
\begin{equation}
3.1\cdot 10^{-12}\lesssim BR_{[J/\psi\rightarrow\ga\ga],max}^{\;a=1,\;x_i=0}\lesssim7.8\cdot 10^{-10}, 
\quad 1.7\cdot 10^{-11}\lesssim BR_{[\Upsilon\rightarrow\ga\ga],max}^{\;a=1,\;x_i=0}\lesssim4.3\cdot 10^{-9},
\label{BRnmNCSM}
\end{equation} 
as was obtained in eqs.~(23,24) from \cite{Melic:2005hb}.

On the other side, the minimal values in the same range of $\Lambda_{\rm NC}$ are
\begin{equation}
2.3\cdot 10^{-13}\lesssim BR_{[J/\psi\rightarrow\ga\ga],min}^{\;a=1,\;x_i=0}\lesssim5.9\cdot 10^{-11}, 
\quad 1.0\cdot 10^{-26}\lesssim BR_{[\Upsilon\rightarrow\ga\ga],min}^{\;a=1,\;x_i=0}\lesssim2.6\cdot 10^{-24}\,.
\label{eq40}
\end{equation} 
By taking  $\vec{E}^2_\th=0,\vec{B}^2_\th=1$ all the results are suppressed by a factor of $3/10$.

 To see the effect of adding to the action the extra pure gauge term associated with 
 the choice $a=3$ in eq.~\eqref{SgC}, corresponding to the deformed nmNCSM with 
 renormalisable pure gauge interactions, we can start from  (\ref{ratef}) 
 and  fix  $a=3$ and $x_4=x_5=x_{11}=0$. The resulting expression is more complicated than \eqref{pdwr},
\begin{align}
\nonumber &\frac{\Gamma^{a=3}(\overline{q}q_{1} \to \gamma \gamma )}
{\Gamma(\overline{q}q_{1} \to \ell^+ \ell^-)}=\frac{M^4}{48 e_q^2\Lambda_{\rm NC}^4(M^2-M_Z^2)^2}
(R \vec{B}_\th^2+S \vec{E}_\th^2),\\
\nonumber&R=3 \left(M^2-M_Z^2\right){}^2 {e_q}^4+36 \sin ^2\left(2 \theta _W\right) 
K_{{\ga\ga\ga}}^2 \left(M^2-M_Z^2\right){}^2 {e_q}^2+20 c_V^q M^2 K_{{Z\ga\ga}} \left(M^2-M_Z^2\right) {e_q}^2\\
&\phantom{R=}-4 \sin \left(2 \theta _W\right) K_{{\ga\ga\ga}} \left(M^2-M_Z^2\right) 
\left(5 \left(M^2-M_Z^2\right) {e_q}^2+18 c_V^q M^2 K_{{Z\ga\ga}}\right) {e_q}+36 {c_V^q}^2 M^4 K_{{Z\ga\ga}}^2\\
\nonumber&S=7 \left(M^2-M_Z^2\right){}^2 {e_q}^4+36 \sin ^2\left(2 \theta _W\right) 
K_{{\ga\ga\ga}}^2 \left(M^2-M_Z^2\right){}^2 {e_q}^2-20 {c_V^q} M^2 K_{{Z\ga\ga}} \left(M^2-M_Z^2\right) {e_q}^2\\
\nonumber&\phantom{S=}+4 \sin \left(2 \theta _W\right) K_{{\ga\ga\ga}} 
\left(M^2-M_Z^2\right) \left(5 {e_q}^2 \left(M^2-M_Z^2\right)-18 {c_V^q} M^2 K_{{Z\ga\ga}}\right) 
{e_q}+36 {c_V^q}^2 M^4 K_{{Z\ga\ga}}^2\,.
\label{eq41}
\end{align}
The maximal values in the allowed regions for $K_{Z\ga\ga}$ and $K_{\ga\ga\ga}$ of the branching ratios, for 
$0.25 \text{  TeV}\leq\Lambda_{\rm NC}\leq 1 \text{  TeV}$, $\vec{E}_\th^2=\vec{B}_\th^2=1$ become
\begin{equation}
2.4\cdot 10^{-12}\lesssim BR_{[J/\psi\rightarrow\ga\ga],max}^{\;a=3,\;x_i=0}\lesssim6.3\cdot 10^{-10}, 
\quad 7.5\cdot 10^{-11}\lesssim BR_{[\Upsilon\rightarrow\ga\ga],max}^{\;a=3,\;x_i=0}\lesssim1.9\cdot 10^{-8}\,,
\label{eq42}
\end{equation}
whereas for the minimal values we get
\begin{equation}
5.1\cdot 10^{-13}\lesssim BR_{[J/\psi\rightarrow\ga\ga],min}^{\;a=3,\;x_i=0}\lesssim1.3\cdot 10^{-10}, 
\quad 4.6\cdot 10^{-12}\lesssim BR_{[\Upsilon\rightarrow\ga\ga],min}^{\;a=3,\;x_i=0}\lesssim1.2\cdot 10^{-9}\,.
\label{eq43}
\end{equation}
They remain more or less in the same order of magnitude as those of eq.~\eqref{BRnmNCSM}, 
safe for the minimal values of the $\Upsilon$ branching ratio, which are hugely increased. 
In particular, the minimal values are always above those of the mNCSM of eq.~\eqref{BRmNCSM}.
In the case $\vec{E}_\th^2=0,\vec{B}_\th^2=1$ the rates are again suppressed but in a different way:
\begin{align}
\nonumber 1.9\cdot 10^{-12}\lesssim BR_{[J/\psi\rightarrow\ga\ga],max}^{\;a=3,\;x_i=0}\lesssim4.8\cdot 10^{-10}, 
\quad 2.3\cdot 10^{-11}\lesssim BR_{[\Upsilon\rightarrow\ga\ga],max}^{\;a=3,\;x_i=0}\lesssim5.9\cdot 10^{-9},\\
3.5\cdot 10^{-14}\lesssim BR_{[J/\psi\rightarrow\ga\ga],min}^{\;a=3,\;x_i=0}\lesssim9.0\cdot 10^{-12}, 
\quad 1.0\cdot 10^{-13}\lesssim BR_{[\Upsilon\rightarrow\ga\ga],min}^{\;a=3,\;x_i=0}\lesssim2.6\cdot 10^{-11}\,.
\label{eq44}
\end{align}
In particular, the minimal values are no longer above the ones corresponding 
to the mNCSM with  $\vec{E}_\th^2=0,\vec{B}_\th^2=1$, but there is still a dramatic increase in 
the minimal possible values for the $\Upsilon$ branching ratio.

So far, the effect of the extra pure gauge term in the deformed nmNCSM with respect to 
the undeformed version is essentially an enhancement of the minimal allowed values of 
the $\Upsilon$ branching ratio.

Next we move on to see the effect of the extra terms in the fermionic action depending on $x_4,x_5,x_{11}$, 
when these parameters take natural values between -1 and 1. 
We set to calculate the maximal and minimal values that the branching ratios obtained from eq.~\eqref{ratef} 
can have for $x_i\in\{-1,1\}$ within the allowed values of $K_{Z\ga\ga}$ and $K_{\ga\ga\ga}$.

In the case $\vec{E}_\th^2=\vec{B}_\th^2=1$, we obtain the following results, as before for scales 
$0.25 \text{  TeV}\leq\Lambda_{\rm NC}\leq 1 \text{  TeV}$:
\begin{align}
&\nonumber 4.5\cdot 10^{-11}\lesssim BR_{[J/\psi\rightarrow\ga\ga],max}^{\;a=3,\;|x_i|\sim 1}\lesssim1.2\cdot 10^{-8}, 
\quad 5.6\cdot 10^{-10}\lesssim BR_{[\Upsilon\rightarrow\ga\ga],max}^{\;a=3,\;|x_i|\sim 1}\lesssim1.4\cdot 10^{-7},\\
&BR_{[J/\psi\rightarrow\ga\ga],min}^{\;a=3,\;|x_i|\sim 1}\sim0, \quad
 BR_{[\Upsilon\rightarrow\ga\ga],min}^{\;a=3,\;|x_i|\sim 1}\sim0.
\label{NCBRx}
\end{align}

\begin{figure}[h]
\begin{center}
  \includegraphics[scale=1]{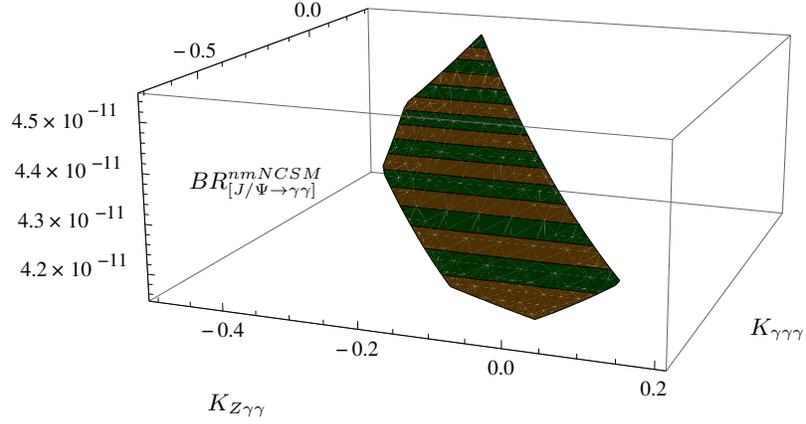}
 \end{center}\vskip-1cm
\caption
{$J/\Psi\rightarrow\ga\ga$ branching ratio as a function of 
$K_{Z\ga\ga}$ and $K_{\ga\ga\ga}$, for $\vec{E}_\th^2,\vec{B}_\th^2\sim 1$, $x_4=x_5=-x_{11}=-1$, 
at the scale $\Lambda_{\rm NC}=1\text{ TeV}$}.
\label{f:3}
\end{figure}

\begin{figure}[h]
 \begin{center}
  \includegraphics[scale=1]{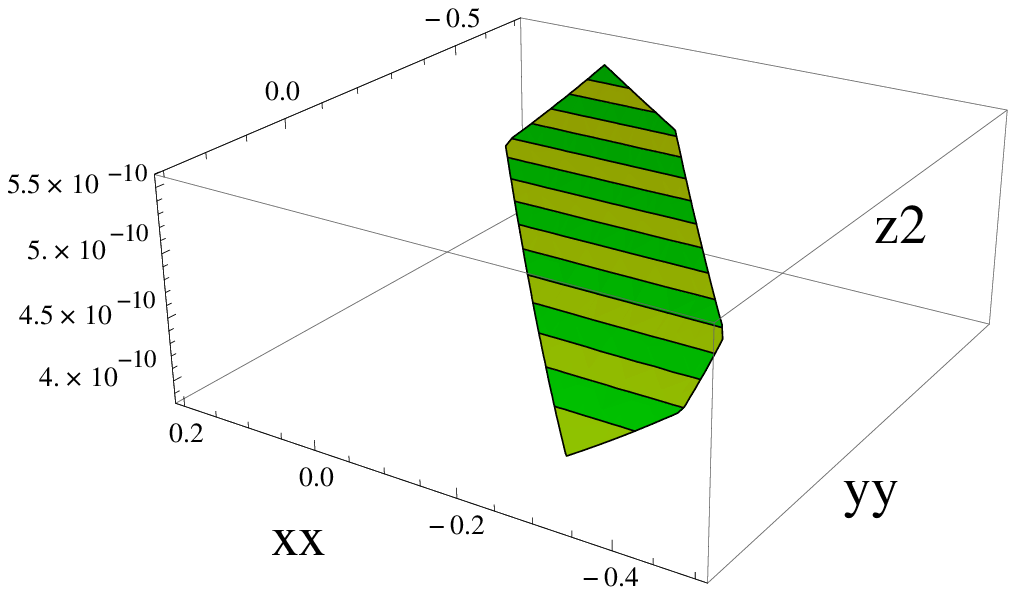}
 \end{center}
\caption
{$\Upsilon\rightarrow\ga\ga$ branching ratio as a function of  
$K_{Z\ga\ga}$ and $K_{\ga\ga\ga}$, for $\vec{E}_\th^2,\vec{B}_\th^2\sim 1$, $x_4=x_5=-x_{11}=-1$, 
at the scale $\Lambda_{\rm NC}=1\text{ TeV}$}.
\label{f:4}
\end{figure}

\begin{figure}[h]
 \begin{center}
  \includegraphics[scale=1]{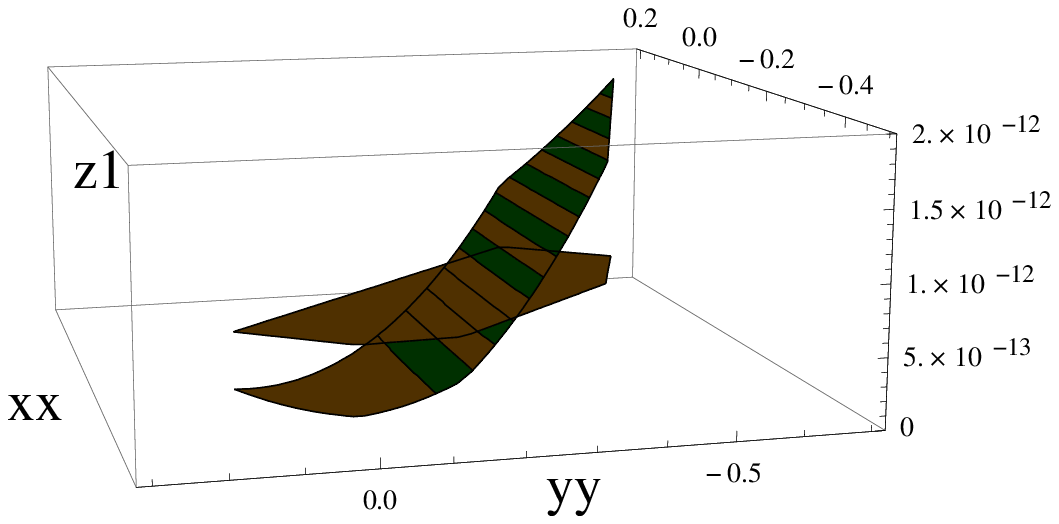}
 \end{center}\vskip-1cm
\caption
{$J/\Psi\rightarrow\ga\ga$ branching ratio as a function of  
$K_{Z\ga\ga}$ and $K_{\ga\ga\ga}$, 
for $\vec{E}_\th^2,\vec{B}_\th^2\sim 1$, $x_4=0.25,x_5=-2.85\cdot 10^{-7},x_{11}=2.86\cdot 10^{-7}$, 
at the scale $\Lambda_{\rm NC}=1\text{ TeV}$. 
The horizontal plane represents the mNCSM branching ratio of $5.1\cdot 10^{-13}$.}
\label{f:5}
\end{figure}

\begin{figure}[h]
 \begin{center}
  \includegraphics[scale=1]{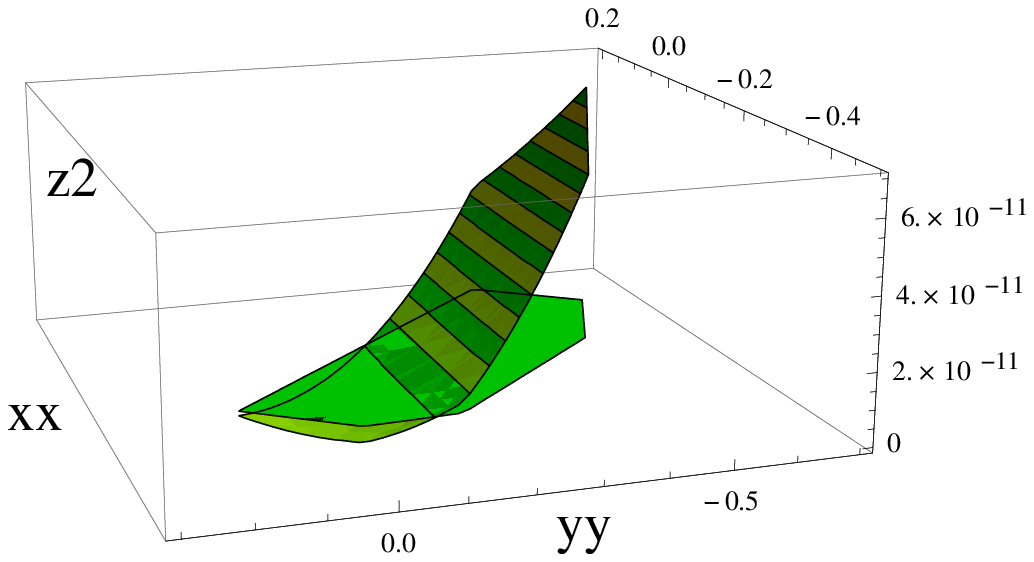}
 \end{center}\vskip-1cm
\caption
{$\Upsilon\rightarrow\ga\ga$ branching ratio as a function of  
$K_{Z\ga\ga}$ and $K_{\ga\ga\ga}$, 
for $\vec{E}_\th^2,\vec{B}_\th^2\sim 1$, $x_4=0.25,x_5=-2.85\cdot 10^{-7},x_{11}=2.86\cdot 10^{-7}$, 
at the scale $\Lambda_{\rm NC}=1\text{ TeV}$. The horizontal plane represents 
the mNCSM branching ratio of $4.6\cdot 10^{-12}$.}
\label{f:6}
\end{figure}

As an explanation of the previous values, for example the first expression $4.5\cdot 10^{-11}\lesssim 
BR_{[J/\psi\rightarrow\ga\ga],max}^{\;a=3,\;|x_i|\sim 1}\lesssim1.2\cdot 10^{-8}$ means 
that the maximum of the $J/\psi$ 
 branching ratio, for the range of scales $0.25 \text{  TeV}\leq\Lambda_{\rm NC}\leq 1 \text{  TeV}$, 
 for all $x_i\in\{-1,1\}$ and for all allowed $K_{Z\ga\ga}$ and $K_{\ga\ga\ga}$, varies between 
 $4.5\cdot 10^{-11}$ --reached at the scale $\Lambda_{\rm NC}=1 \text{ TeV}$, and $1.2\cdot 10^{-8}$, 
 reached at $\Lambda_{\rm NC}=0.25 \text{ TeV}$. Note that the dependency of the rates on $\Lambda_{\rm NC}$ 
 factorises, as seen in eq.~\eqref{ratef}. Each value is reached for a particular value of 
 $x_i,K_{Z\ga\ga},K_{\ga\ga\ga}$: 
\begin {align} &\text{Maxima}\label{max}\\
\nonumber&\phantom{J\psi}J/\psi: x_4=-1.00,x_5=-1.00,x_{11}=1.00,K_{Z\ga\ga}=-0.254,K_{\ga\ga\ga}=0.129,\\
\nonumber&\phantom{J\psi}\Upsilon:x_4=-1.00,x_5=-1.00,x_{11}=1.00,K_{Z\ga\ga}=0.00950,K_{\ga\ga\ga}=-0.576,
\end{align}
whereas for the minima the numerical results are not reliable due to precision issues.
Figures \ref{f:3} through \ref{f:6} show the resulting branching ratios as functions of 
$K_{Z\ga\ga}$ and $K_{\ga\ga\ga}$ for the particular values of the $x_i$ that yielded
the above maxima and (approximate) minima, respectively, at the scale $\Lambda_{\rm NC}=1 \text{ TeV}$. 
The maxima of figures \ref{f:3} and \ref{f:4} correspond to the values appearing at 
the left of the inequalities in eq.~\eqref{NCBRx}. The minima of figures \ref{f:5} and \ref{f:6} are practically zero.


For completeness we show the corresponding results in the $\vec{E}_\th^2=0,\;\vec{B}^2_\th=1$ case:
\begin{align}
&\nonumber2.0\cdot 10^{-11}\lesssim BR_{[J/\psi\rightarrow\ga\ga],max}^{\;a=3,\;|x_i|\sim 1}\lesssim5.1\cdot 10^{-9}, 
\quad 2.0\cdot 10^{-10}\lesssim BR_{[\Upsilon\rightarrow\ga\ga],max}^{\;a=3,\;|x_i|\sim 1}\lesssim5.0\cdot 10^{-8},\\
&BR_{[J/\psi\rightarrow\ga\ga],min}^{\;a=3,\;|x_i|\sim 1}\sim 0, 
\quad  BR_{[\Upsilon\rightarrow\ga\ga],min}^{\;a=3,\;|x_i|\sim 1}\sim 0,
\label{eq46}
\end{align}

The previous values are reached for the following values of the parameters:
\begin {align} &\text{Maxima}\label{max2}\\
\nonumber&\phantom{J\psi}J/\psi: x_4=-1.00,x_5=-1.00,x_{11}=-1.00,K_{Z\ga\ga}=0.00950,K_{\ga\ga\ga}=-0.576,\\
\nonumber&\phantom{J\psi}\Upsilon:x_4=1.00,x_5=1.00,x_{11}=1.00,K_{Z\ga\ga}=0.00950,K_{\ga\ga\ga}=-0.576,
\end{align}
where again we do not display the position of the minima since our numerical results are not reliable.
The effect of the $x_i$ terms is clearly to allow for much larger maximum values 
and much lower minimum values of the branching ratios with respect to 
the results in the undeformed nmNCSM or the mNCSM. The maximum values are  
increased up to 2 orders of magnitude with respect to the mNCSM result of 
eq.~\eqref{BRmNCSM} and one order of magnitude with respect to the results 
in eq.~\eqref{BRnmNCSM} corresponding to the undeformed version of the nmNCSM.

Despite the fact that the allowed values for the minima experience an important decrease, 
for typical values of the parameters $K_{Z\ga\ga},K_{\ga\ga\ga},x_4,x_5,x_{11}$ 
the rates are enhanced with respect to the mNCSM result. To justify this claim 
we have computed the branching ratios for random values of the above parameters in 
the allowed region for $K_{Z\ga\ga},K_{\ga\ga\ga}$ displayed in figure \ref{f:0} and for $x_i\in\{-1,1\}$. 
For up to one million configurations of these parameters, at the scale $\Lambda_{\rm NC}=1 \text{ TeV}$, 
we found that, for $\vec{E}_\th^2=\vec{B}^2_\th=1,\; (\vec{E}_\th^2=0,\;\vec{B}^2_\th=1)$, 96(88)\% of 
the configurations yield $J/\psi$ branching ratios
larger than their corresponding values in the mNCSM, whereas in the $\Upsilon$ case 
the percentages are 97(89)\%. The percentage of configurations yielding a $10\times$ 
increase over the mNCSM values are 44(42)\% in the $J/\psi$ case and 55(40)\% in the $\Upsilon$ case. 
There is a  $50\times$ increase for 2(4)\% of the $J/\psi$ 
configurations and for 4(3)\% of the $\Upsilon$ configurations.

We have also estimated the portion of the parameter space which, 
at the scale $\Lambda_{\rm NC}=1 \text{ TeV}$, yields values of 
the branching ratios which are larger than the maximum values in 
the undeformed version of the nmNCSM, which are given in eq.~\eqref{BRnmNCSM}. 
The percentage of configurations that satisfy this requirement for 
$\vec{E}_\th^2=\vec{B}^2_\th=1,\; (\vec{E}_\th^2=0,\;\vec{B}^2_\th=1)$, 
is given in the $J/\psi$ case by 62(55)\%, 
whereas in the $\Upsilon$ case it is equal to 84(67)\%.

To complete our numerical estimates, we should consider the case when the parameters $x_i$ 
take values that are necessarily non-natural, i.e, not necessarily restricted to be between -1 and 1. 
An analysis of the dependence of the branching ratios on the parameters $K_{Z\ga\ga},\,K_{\ga\ga\ga},\,x_i$ 
shows that, independently of the $x_i$, the branching ratios as a function of 
$K_{Z\gamma\gamma}$ and $K_{\gamma\gamma\gamma}$ define a concave parabolic surface (in the $J/\Psi$ case) 
or a hyperbolic surface ($\Upsilon$ case). Thus the maxima always appear in the boundary of the region of 
the parameter space, as has happened with our  results in eqs.~\eqref{max} and \eqref{max2}. 
It is clear that allowing a wider range of the $x_i$ will directly yield greater maxima; 
since the dependence of the branching ratios is quadratic in the $x_i$, 
we expect that an increase of an order of magnitude in the range of the $x_i$ yields 
a two orders of magnitude increase in the maximum values of the branching ratios. 
By doing some numerical computations this seems to be roughly the case (modulo a factor between 0.45 and 0.77). 
In  particular, taking $x_i\in\{-100,100\}$, we get, for  $\vec{E}^2_\th=\vec{B}^2_\th=1$ and for scales 
$0.25 \text{  TeV}\leq\Lambda_{\rm NC}\leq 1 \text{  TeV}$:
\begin{align}
& 3.5\cdot 10^{-7}\lesssim BR_{[J/\psi\rightarrow\ga\ga],max}^{\;a=3,\;|x_i|\sim100}\lesssim9.4\cdot 10^{-5}, 
\quad 3.2\cdot 10^{-6}\lesssim BR_{[\Upsilon\rightarrow\ga\ga],max}^{\;a=3,\;|x_i|\sim100}\lesssim8.1\cdot 10^{-4}.
\label{max3}
\end{align}
The values at the scale $\Lambda=0.25 \text{ TeV}$ are within the experimental bounds: 
$BR(J/\psi \rightarrow \gamma \gamma)< 2.2 \times 10^{-5}$ \cite{Amsler:2008zz} 
and $BR(\Upsilon \rightarrow \gamma \gamma) \lesssim 10^{-4}$ \cite{Brambilla:2004wf}.

\section{Discussion and conclusions}

In this paper we have introduced a deformed fermion lagrangian for the nmNCSM 
and we have applied it to the phenomenological estimate of 
the quarkonia decay rates into two photons.

First of all, let us recall that the non-minimal version of the NCSM gauge sector
includes traces over the representations of all the massive particle multiplets with different 
quantum numbers that appear in the total lagrangian of the model with covariant derivatives acting on them, 
i.e., with terms of the type
$i\overline{\widehat\psi}\star \hat\Dirac\star\widehat\psi,\;
(\hat D^\m\widehat\Phi)^{\dagger} \star(\hat D_\m\widehat\Phi),$ 
\cite{Behr:2002wx,Melic:2005fm}.

Second, we have argued why the use of the nmNCSM, and in particular its deformed version, 
should be favoured over the mNCSM; this is because of the proven renormalisability of 
the pure gauge sector and the existing hints that this property might not be spoilt when 
the effects of matter loops are taken into account. 
The motivations for introducing deformation terms in the fermion lagrangian come essentially 
from our lack of knowledge of the renormalisability properties of the matter sector of the full nmNCSM, 
and also from the fact that, for the gauge interactions, renormalisability --and even finiteness for 
the first noncommutative corrections-- was only achieved by adding 
a deformation term to the starting lagrangian. 
Thus, our deformed fermion lagrangian should be taken as an effective lagrangian for 
the nmNCSM as long as the renormalisability properties are unknown. From the effective theory point of view, 
the values of the $x_i$ parameters could be constrained by experimental measurements. 

It should be noted that the photon polarisation is known to be modified by noncommutativity, which
causes vacuum birefringence \cite{AlvarezGaume:2003mb}, at least in the standard approach to noncommutative theories. This follows from computations of the one-loop photon self-energy. It should be interesting to analyse this issue in the enveloping algebra approach. Of course our new fermion-photon interactions, coming from
the $x_i$-dependent terms, will also affect the photon polarisation at one-loop, 
which could be used for further experimental tests. 
However this is not straightforward since, according to ref.~\cite{Bichl:2001cq}, the photon two point function, due to gauge invariance, 
will only be modified at order $\theta^2$ and beyond, so that a consistent computation would 
imply the use of SW maps up to O($\theta^2$).

Aside from possible experimental measurements, there is hope that further 
investigations about renormalisability properties could impose constraints 
on the free deformation parameters $x_i$; it could happen that some or all of them were fixed uniquely, 
as happened for the gauge sector of the nmNCSM, and in that case 
the new fermion sector would not be effective but part of a theory well defined in the ultraviolet.

In deriving the deformed fermion lagrangian, we did not consider terms that altered 
the tree-level fermion propagator. This was done because, on the one hand, 
we should expect noncommutative effects to appear as weak quantum corrections, 
and thus we do not find desirable to break the usual 
Lorentz-invariance matter dispersion relation $p^2=m^2$ at tree level. 
On the other hand, we wanted to apply the framework to the computation of decay rates, 
which implies the calculation of S-matrix elements. In ordinary space-time one uses the LSZ formula, 
which relies on general properties of the pole structure of the Green functions of the theory 
(and in particular the 2-point function) which follow only from Poincar\'e invariance. 
In noncommutative space-time the usual Lorentz invariance does not hold, 
and thus a proper all-order definition of S-matrix elements seems challenging 
(in the case of noncommutative theories that do not make use of Seiberg-Witten maps, 
some advances have been done in this respect, see for example ref.~\cite{Grosse:2008dk}). 
Nevertheless, since we were computing S-matrix elements at tree level 
and we were not altering the tree-level 2-point function, we believe that 
the usual LSZ formalism should hold at this level. As a consistency check, 
our S-matrix amplitudes satisfy the usual Ward identities associated with U(1) gauge invariance, 
which are derived in ordinary space-time as a consequence of gauge symmetry 
and the pole structure of the diagrams. Nevertheless, 
a deeper understanding of the S-matrix and the LSZ formalism 
in noncommutative theories is still needed. 
We also recall that some of the terms considered in eq.~\eqref{fH} violate CPT exclusively in the weak sector; 
also, there appear C,T violations in the strong and hypercharge sectors. 
This could be of phenomenological interest for searches of Physics beyond the Standard Model; 
note that the violations are very small since they appear purely as noncommutative effects.

Concerning our results for the quarkonia decay amplitudes, it should be noticed first that the on-shell amplitudes
turned out to be independent of the C, T violating terms $t_8,t_{10}$; 
also, though they are dependent on $x_{11}$, 
which is associated to CPT violations in the weak sector, 
no CPT violating interactions contributed to the result. Ref.~\cite{Grimstrup:2002af} 
argued that to get one-loop renormalisability in noncommutative QED, the term $t_7$ of eq.~\eqref{fH} 
should be added to the bare lagrangian. This does not conflict with the fact that the quarkonia decay 
amplitudes are not apparently influenced by the terms with three $\ga^\m$ matrices in eq.~\eqref{fH}: 
first, we absorbed the contributions of $t_7$ in the terms $t_4-t_6,t_8,t_{10},t_{11}$ by using field redefinitions, 
and, though the on-shell amplitudes did not depend on the parameters $x_6,x_8,x_{10}$, 
the amplitudes evaluated at arbitrary momenta did. 
With respet to the numerical results obtained for the decay rates, we showed that the effect 
of the extra gauge term in the action of the extended nmNCSM is essentially 
to raise the minimum allowed values of the $\Upsilon\rightarrow\ga\ga$ decay rate, 
masking the destructive contributions that were found in the undeformed nmNCSM in ref.~\cite{Melic:2005hb}. 
However, when the new terms of our deformed fermion lagrangian are taken into account, this effect disappears, 
and both the minimum and maximum allowed values for the decay rates experience high decreases and increases, 
respectively.

Having very low allowed values for the quarkonia decay rates is no good news for 
the possibility of comparison with experiments in order to confirm or falsify 
the theoretical predictions: in principle, non-zero  $q\bar q_1\rightarrow\ga\ga$ 
decay rates could be taken as a signal of noncommutativity, but if the models 
allow for extremely small values it would be  difficult to discard them. 
However, the panorama 
is more promising because, despite the minimum possible values are very small, 
for most configurations of the parameters (between 88 and 97 \% at $\Lambda_{\rm NC}$=1 TeV 
for natural values of the parameters) we obtain branching ratios that are greater than the ones 
that were computed for the mNCSM. Furthermore, for a big portion of 
the parameter space (between 55 and 84 \% under the same conditions), 
the decay rates, computed from deformed nmNCSM, are actually larger than the maximum values 
that were found in the case of the undeformed nmNCSM.

Thus, in general we get constructive contributions to the decay rates 
and the model allows for greater values of the branching ratios that 
the ones that had been previously found; the maximum values increase by up to two orders of magnitude.
Possible future studies of the renormalisability of the matter sector could help 
to restrict the allowed values in the parameter space,
as happens in the gauge sector with the ambiguity parameter $a$ of 
eq.~\eqref{SgC} forced to be equal to 3, and this could make the model more predictive or falsifiable. 
Again, we recall the result of ref.~\cite{Grimstrup:2002af}, which in the QED case argues 
that one-loop renormalisability demands to add only the term $t_7$ of eq.~\eqref{fH} to the lagrangian; 
this may also happen in the NCSM but the result cannot be directly extrapolated since, 
after the expansion with the SW map, the NCSM lagrangian is not given by a sum of lagrangians for 
the different gauge groups due to the appearance of interactions between the different gauge fields.

Todays existing experimental limit for 
the branching ratio of the $J/\psi \rightarrow \gamma \gamma$ decay can be found in 
``Review of particle physics,'' under the C symmetry violating modes, and is 
$BR(J/\psi \rightarrow \gamma \gamma)< 2.2 \times 10^{-5}$
\cite{Amsler:2008zz}. 

With respect to the $\Upsilon$ case, as it was commented in ref.~\cite{Melic:2005hb}, the existing limit for 
the branching ratio of the $\Upsilon \rightarrow \gamma \gamma$ decay comes from a very old
CLEO-III experiment \cite{Brambilla:2004wf} and it 
indicates that with present data the detection of $BR(\Upsilon \rightarrow \gamma \gamma)$
below $ 10^{-4}$ would be hopeless.

As follows from eq.~\eqref{max3}, the previous experimental bounds are reachable in our model at 
the scale $\Lambda=0.25 \text{TeV}$ for values of $x_i$ tuned to be around 100, 
and for lower scales for higher values of the $x_i$. Thus there is some hope that 
the phenomenology of our model could be relevant.

Furthermore, despite the fact that all experiments including hadrons are extremely hard to perform and analyse
due to the  huge background signals,
the large number of heavy quark-antiquark pairs harvested at LHCb; i.e. 
$~10^{12}$ $B\bar B$ pairs per year \cite{Klous:2005vp},
and probably ($10^{14}$ $D\bar D$; $~10^{18}$ $K\bar K$), give us hope that 
experimental branching ratios $BR(\Upsilon \to \gamma\gamma)\sim 10^{-9}$
and $BR(J/\psi \to \gamma\gamma)\sim 10^{-11}$  could be accessible,
thus reaching our maximum predicted values for the rates for $|x_i|\sim1$ (\ref{eq42}-\ref{eq46}). 
Certainly these experiments would produce at least  reliable (and much lower) bounds;
perhaps they could even measure these processes, depending on the scale of noncommutativity.
We hope that the importance of a possible
discovery of space-time non-commutativity will convince
experimentalists to look for SM forbidden decays in hadronic physics.

\subsection*{Acknowledgments}
The authors wish to thank Maja Buric, Harald Grosse and 
Carmelo P. Mart\'{\i}n for their useful comments.
C.~T. wishes to thank the members of the Theoretical Physics Department at 
the Rudjer Bo\v{s}kovi\'{c} Institute, Zagreb, for their kind hospitality; 
his work received financial support from  MICINN through grant FIS2005-02309.
The work of J.~T. is supported by 
the Croatian Ministry of Science Education and Sports project No. 098-0982930-2900,
and is in part supported by the EU (HEPTOOLS) project under contract MRTN-CT-2006-035505.

\end{document}